\documentclass[journal,10pt]{IEEEtran}

\usepackage{subfigure}
\usepackage{epsfig,graphicx,rotating}
\usepackage{epstopdf}

\usepackage{amssymb,amsmath}
\usepackage{tikz} 
\usepackage{tikz-3dplot}
\usepackage{color}
\usepackage{cite}      
\usepackage{afterpage}
\usepackage{hyperref}
\usepackage{ellipsis}
\usepackage{multirow}
\usepackage{amsmath,mathtools}
\usepackage{hvfloat}
\usepackage{tabu}
\usepackage{soul}

\newcommand{\norm}[1]{\left\lVert#1\right\rVert}

\makeatletter
\newcommand{\vast}{\bBigg@{4}}
\newcommand{\Vast}{\bBigg@{5}}
\makeatother

\usepackage{color}
\usetikzlibrary{decorations.pathreplacing,decorations.pathmorphing,decorations.markings,shapes,backgrounds,patterns,decorations.text,shapes.callouts,calc,arrows,positioning}
\tikzset{  zigzag/.style={to path={ -- ($(\tikztostart)!.55!-7:(\tikztotarget)$) -- ($(\tikztostart)!.45!7:(\tikztotarget)$) -- (\tikztotarget) \tikztonodes}}}
\tdplotsetmaincoords{0}{0} 
\tdplotsetrotatedcoords{0}{0}{0} 

\tikzset{
  pics/lightning/.style 2 args={code={
    \draw [thick, arrows={-Stealth[scale=2]}] (#1) -- 
               ($(#1)!.5!(#2) + (.25,-.25)$) -- 
               ($(#1)!.5!(#2) + (-.25,.25)$) --
               (#2);
  }}}

\pgfdeclaredecoration{lightning bolt}{draw}{
\state{draw}[width=\pgfdecoratedpathlength]{
  \pgfpathmoveto{\pgfpointorigin}%
  \pgfpathlineto{\pgfpoint{\pgfdecoratedpathlength*0.6}%
    {-\pgfdecoratedpathlength*.1}}%
  \pgfpathlineto{\pgfpoint{\pgfdecoratedpathlength*0.55}{0pt}}%
  \pgfpathlineto{\pgfpoint{\pgfdecoratedpathlength}{0pt}}%
  \pgfpathlineto{\pgfpoint{\pgfdecoratedpathlength*0.4}%
    {\pgfdecoratedpathlength*.1}}%
  \pgfpathlineto{\pgfpoint{\pgfdecoratedpathlength*0.45}{0pt}}%
  \pgfpathclose%
}%
}
	
\tikzset{
  htree leaves/.initial=2,
  sibling angle/.initial=20,
  htree level/.initial={}
}



\begin{document}

\title{
Massive MIMO for Connectivity with Drones: Case Studies and Future Directions}

\author{Prabhu~Chandhar$^1$, Member, \textit{IEEE} and Erik~G.~Larsson$^2$, Fellow, \textit{IEEE}\\
$^1$Chandhar Research Labs, Chennai, 600 030 Tamil Nadu, India\\
$^2$Dept. of Electrical Eng. (ISY), Link\"{o}ping University, 581 83 Link\"{o}ping, Sweden\\
Email: prabhu@chandhar-labs.com, erik.g.larsson@liu.se
  \thanks{This work was supported by the Swedish Research Council (VR), ELLIIT, and Security-Link. Parts of this work were performed when the
first author was with Link\"{o}ping University.}
}

\maketitle

\begin{abstract}\label{abstract}
Unmanned aerial vehicles (UAVs), also known as drones, are proliferating. 
Applications such as surveillance, disaster management, and drone racing place high requirements on the communication with the drones in terms of throughout, reliability and latency. 
Existing wireless technologies, notably WiFi, that are currently used for drone connectivity are limited to short ranges and low-mobility situations. 
A new, scalable technology is needed to meet future demands on long connectivity ranges, support for fast-moving drones, and the possibility to simultaneously communicate with entire swarms of drones. 
Massive multiple-input and multiple-output (MIMO), a main technology component of emerging 5G standards, has the potential to meet these requirements. 
\end{abstract}


\IEEEpeerreviewmaketitle

\section{\textbf{Drone Technology is Proliferating}}
Traditionally, unmanned aerial vehicles (UAVs) in the form of drones have been predominantly used for military applications such as battlefield and airspace surveillance, and border patrol. 
More recently, due to technological advancements in battery and control technology, and miniaturization of electronics, the use of drones for civilian applications is exploding. 
Such applications include traffic and crowd monitoring, wildlife conservation, search-and-rescue operations during natural  disasters, inspection of non-reachable areas, and the transportation of goods \cite{Hayat2016,gupta2016,zeng2016,raja2017,mozaffari2018}. 
Inexpensive lightweight   quadcopter drones are widely available and routinely used for  aerial photography and drone racing. 
More applications are likely to emerge \cite{asadpour2014,mozaffari2018,zeng2016,raja2017,ieee_usa}. 
Forecasts show that by 2022, the global market for drone-based business services will be over USD 100 billion per year \cite{goldmansachs}.  

Government bodies and standardization organizations across the world have taken initiatives towards spectrum allocation and safe drone operations   \cite{kerczewski2013,ieee_usa,easa,3gpp_uav_report}. 
For example, in 2014, the United Nations Office
for the Coordination of the Humanitarian Affairs (OCHA) released an important policy document
on the use of civilian UAVs in humanitarian settings \cite{OCHA}.
Several 5G use cases for drones were identified by 3GPP \cite{3GPP-QLCM,metis2015,3gpp_uav_report}: broadband access to under-developed areas, hotspot coverage during sport events, and rapidly deployable ``on-demand'' densification. 
Field trials have been carried out in cellular networks by major telecom companies including AT\&T, Qualcomm, Nokia and China Telecom,   Ericsson and China Mobile \cite{att_flying_cow,lin2018,intel,ericsson_report}. 
Intel entered into the UAV market by introducing MODEM for drones \cite{intel}. 
In cellular communications, the uses of drones is twofold. 
On the one hand, the drones can act as flying base stations (BSs) to provide hot-spot coverage during emergency situations \cite{att_flying_cow}. 
On the other hand,   drones acting as user equipments (UEs) can be served by  ground stations (GSs) or existing cellular BSs \cite{bulut2018}. 
In this paper, we focus on  applications and use cases in which the drones act as UEs. 

The potential of multi-drone networks is particularly significant as many small drones can be deployed in a short time to cover a large geographical area \cite{olsson2010,andre2014,OCHA,Northfield2013,gupta2016,Jaber_UAV,mademlis2019,Campion2018,diwu2017}. 
In many applications, the drones, acting as UEs, would stream high-quality imagery to a GS \cite{wang2017_spst,Northfield2013,olsson2010}. 
As the required throughputs are very high, existing wireless technologies are incapable of providing adequate service. 
In this article, we make the case that massive multiple-input and multiple-output (MIMO) \cite{Marzetta16Book}, a main 5G physical-layer technology component, can enable reliable high-rate communication with swarms of drones simultaneously. 
Note that only the GS would be equipped with a massive MIMO antenna array, and that each drone would have a single antenna, 
enabling light-weight drones. 
 
\section{ \textbf{Drone communication requirements and challenges}}

\begin{table*}[!htbp]
\caption{Communication requirements for drones in different use cases \cite{Hayat2016,zeng2016}.}\label{table_use_cases}
\tabulinesep=3mm
    \begin{tabu}{ | p{3.1cm}  | p{1.3cm} | p{2.1cm} | p{3.8cm} | p{2.1cm} | p{1.35cm} | p{1.5cm} |}    \hline
    \textbf{Use cases} &  \textbf{Data type} &  \textbf{Data rate per drone} &  \textbf{Number of drones} &  \textbf{Range} &  \textbf{Mobility}&  \textbf{Latency} \\ \hline
    \begin{minipage}[t]{4.5cm}Crowd Surveillance,\\ Event coverage, \\Environment monitoring  \end{minipage} & Video & \begin{minipage}[t]{3.5cm} Hundreds of Mbps\\ (uncompressed)\\ Tens of Mbps\\ (compressed) \end{minipage}&   10 -- 100 depending on the size of the area   & 100 m -- 3 km & 10 -- 20 m/s & 10 -- 100 ms\\ \hline
		Agriculture         & Image and video&  Tens of Mbps & 1 -- 100 depending on the size of the area and, image and video resolution & 5 m -- 3 km & 5 -- 20 m/s & 10 -- 100 ms \\ \hline
		 \begin{minipage}[t]{4.75cm} Disaster management,\\ Search and rescue operation  \end{minipage}  & Video&  \begin{minipage}[t]{3.5cm}Hundreds of Mbps\\ (uncompressed)\\ Tens of Mbps\\ (compressed) \end{minipage}  &10 -- 100 depending on the size of the area and, image and video resolution  &  50 m -- several km & 0 -- 30 m/s &   $<$40 ms\\ \hline
    Aerial videography & Image and video&   \begin{minipage}[t]{3.5cm}Hundreds of Mbps\\ (uncompressed)\\ Tens of Mbps\\ (compressed) \end{minipage}&  1 -- 10 & 100 m -- 3 km & 0 -- 10 m/s & 100 ms\\ \hline	
		Drone racing & Video&   \begin{minipage}[t]{3.5cm}Tens of Mbps \\(uncompressed)\end{minipage}& 50 -- 100  & 50 m -- 1 km & 55 m/s& $<$ 40 ms \\ \hline	

    \end{tabu}
		\end{table*}

		\begin{figure*}[!htbp]
\centering
\includegraphics[scale=.5]{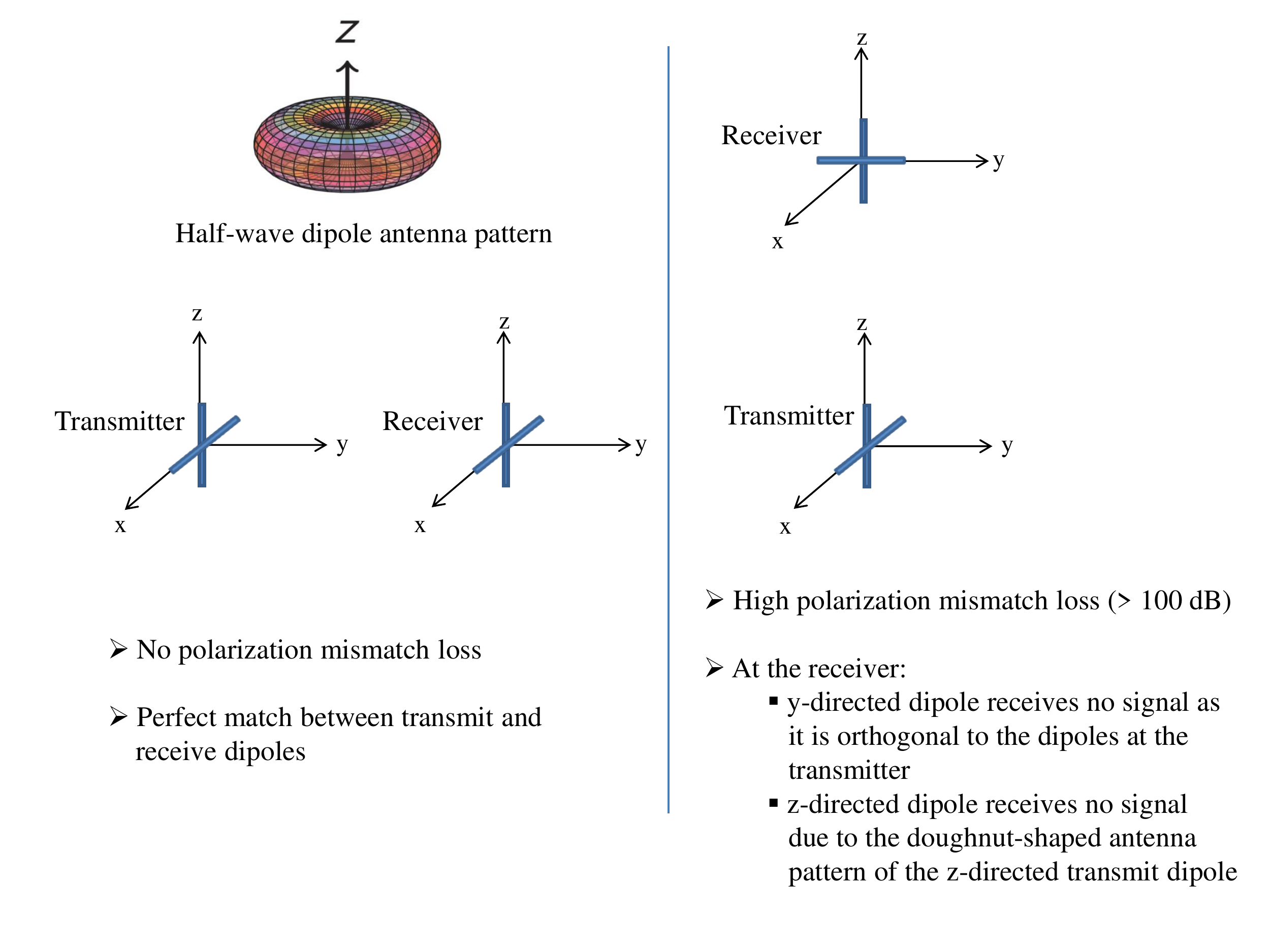}
\caption{Illustration of polarization loss with different transmit and receive antenna orientations in LoS.}
\label{Picture_Pol_Loss}
\end{figure*} 

Table~\ref{table_use_cases} exemplifies the communication requirements of
drone networks for various use cases
\cite{Hayat2016,asadpour2014,zeng2016}. Some general issues are:
\begin{itemize}
\item The channel between the GS and the drones is predominantly line-of-sight (LoS). Since the drones are moving in 3D space, 
due to flight dynamics (roll, pitch, and yaw rotations), the gains and polarizations of the antennas vary with time. 
As a result, maintaining connectivity is challenging \cite{yanmaz2013,chen2018}. 
Note that polarization mismatch is not a major problem in cellular communications, because due to the large number multi-path components in cellular environments, the probability of having a complete mismatch is negligible as the BSs employ dual-polarized antennas. 
In contrast, in drone communications, due to the absence of  significant multi-path propagation, a drone's dynamic movement in 3D space can result in very frequent polarization mismatch events. 
  For example, Figure \ref{Picture_Pol_Loss} illustrates a setup with a fixed transmit antenna (dual-polarized, half-wave dipole) and two different positions (and orientations) of the receiver antenna (dual-polarized, half-wave dipole). It can be noted that even with dual-polarized antennas the loss can be significant in LoS channel conditions (for more details, see \cite{prabhu_TWC}). 
  Therefore, the transmit and the receive antennas must be designed to limit polarization losses.  
   
 \begin{figure*}[!htbp]
\centering
\hspace{-.5cm}\subfigure[Area covered by the image sensor when the drone is at a particular altitude.]{
\tdplotsetmaincoords{70}{160}
\begin{tikzpicture}[scale=.425,line join = round,line cap = round,>=triangle 45,tdplot_main_coords]

 \draw[fill=gray!4] (-5+1,-5,0) --  (-5+1,10,0) -- (5+5,10,0) -- (5+5,-5,0) -- (-5+1,-5,0);

 \draw[-] (-5+1,-5,0) -- node[fill=white,rotate=30,inner sep=-1.25pt,outer sep=0,anchor=center]{$\approx$} (-5+1,10,0);
 \draw[-] (-5+1,10,0) -- node[fill=white,rotate=90,inner sep=-1.25pt,outer sep=0,anchor=center]{$\approx$} (5+5,10,0);
 \draw[-] (5+5,10,0) -- node[fill=white,rotate=30,inner sep=-1.25pt,outer sep=0,anchor=center]{$\approx$} (5+5,-5,0);
 \draw[-] (5+5,-5,0) -- node[fill=white,rotate=90,inner sep=-1.25pt,outer sep=0,anchor=center]{$\approx$} (-5+1,-5,0);

  \coordinate (O) at (0, 0, 0);   
	\coordinate (Y) at (0, 6, 0);   
	\coordinate (Z) at (0, 0, 5);   
	\coordinate (X) at (6, 0, 0);
  \coordinate (M2) at (4.5, 0, 0); 	
	\coordinate (M3) at (4.5, 8, 0); 	
	\coordinate (M4) at (0, 8, 0);   
	\coordinate (M) at (1.5, 2, 9);
  \coordinate (Q1) at (4.5/2+4-1, 8/2, 0); 
	\coordinate (Q2) at (4.5/2+4, 8/2, 0); 
	\def\dx{4.5/4};	
	\def\dy{8/4};	
	\def\L{3};
	\def\H{15};
	\coordinate (R) at (1.5, 2, \H); 	
	\coordinate (T1) at (1.5+\dx/2, 2, \H); 	
	\coordinate (T2) at (1.5/2+\dx/2+4, 2, \H);
  \coordinate (T11) at (1.5+\dx/2, 2, \H-\L); 	
	\coordinate (T21) at (1.5/2+\dx/2+4, 2, \H-\L);
	\coordinate (M41) at (1.5+\dx/2, 2-\dy/2, \H); 	
	\coordinate (M31) at (1.5-\dx/2, 2-\dy/2, \H); 	
	\coordinate (M21) at (1.5-\dx/2, 2+\dy/2, \H);  	
	\coordinate (O1) at (1.5+\dx/2, 2+\dy/2, \H);

  \draw[-,fill=blue!5] (M4) -- (M3) -- (M2) -- (O) -- cycle;

   \foreach \x in {0,0.5,...,4}
        \foreach \y in {0,0.5,...,8}
        {
            \draw[-,thin,gray!30] (\x,0) -- (\x,8);
            \draw[-,thin,gray!30] (0,\y) -- (4.5,\y);
        }

  \draw[thick] (M41) -- (M31) -- (M21) -- (O1) -- cycle;
  \draw[thick] (M4) -- (M3) -- (M2) -- (O) -- cycle;

  \draw[<->,>=stealth] (M2) -- (M4) node [midway,fill=white] {FOV};
  \draw[<->,>=stealth] (Q2) -- (T21)  node [midway,fill=white] {Altitude};
  \draw[<->,>=stealth] (T21) -- (T2)  node [midway,fill=white] {Focal length};

    \draw[<->,>=stealth] (0, 0-2, 0) -- (4.5, 0-2, 0) node [midway,fill=white] {\tiny $r_{px} \!\! \cdot\! \mathrm{GSD}$};
    \draw[<->,>=stealth] (-2, 0, 0) -- (-2, 8, 0) node [midway,fill=white] {\tiny $r_{py} \!\! \cdot\! \mathrm{GSD}$};
    
    	\draw[-] (0, 0-2+.5, 0) -- (0, 0-2-.5, 0);
    	\draw[-] (4.5, 0-2+.5, 0) -- (4.5, 0-2-.5, 0);
    	\draw[-] (-2-.5, 0, 0) -- (-2+.5, 0, 0);
    	\draw[-] (-2-.5, 8, 0) -- (-2+.5, 8, 0);

	\draw[-] (T1) -- (T2);
  \draw[-] (Q1) -- (Q2);
  \draw[-] (T11) -- (T21);

  \draw[dashed] (M4) -- (M41);
  \draw[dashed] (M3) -- (M31);
  \draw[dashed] (M2) -- (M21);
  \draw[dashed] (O) -- (O1);

 \draw[-,fill=blue!15] (M41) -- (M31) -- (M21) -- (O1) -- cycle;

 \draw[-latex,red] (1,11,-.25) node[right,text=red]{Ground image} to[out=180,in=20] (3,6.5,0);
 \draw[-latex,red] (1,11,\H+2) node[right,text=red]{Camera image} to[out=180,in=50] (1.5,2,\H);
 \draw[-latex,red] (7,13,-1) node[right,text=red]{Area to be covered} to[out=170,in=20] (8,9,0);

\draw[-] (4.25, 8+.5, 0) -- (4.25, 8+1, 0); 	
\draw[-] (3.75, 8+.5, 0) -- (3.75, 8+1, 0); 	
\draw[->,>=stealth] (4.75, 8+.75, 0) -- (4.25, 8+.75, 0); 


\draw[<->,>=stealth] (1.85,0,8) node[below,text=red]{} to[out=330,in=220] (0,.3,8); 	
	
	\node at (1.85,2.5,7.95) {AOV};
	
\draw[<->,>=stealth] (M2) -- (M4) node [midway,fill=white] {FOV};

 \draw[-latex,red] (4,12,-.5) node[right,text=red]{GSD} to[out=180,in=320] (4.1, 8+.75, 0);
 
 \node at (8,6,0) {$A_\mathrm{drone}$};
  \node at (1.5,1.25) {\small $ A_\mathrm{image}$};

\draw[-,-latex] (9-3,-10,0) -- (9-3,-10,0+2) node[anchor=north east]{\scriptsize $z$};
\draw[-,-latex] (9-3,-10,0) -- (9-3,-10-4,0) node[anchor=north east]{\scriptsize $y$};
\draw[-,-latex] (9-3,-10,0) -- (9-3-2,-10,0) node[anchor=north east]{\scriptsize $x$};
  
\end{tikzpicture}
\label{GSD_Focal_Length_Illustration}}
\subfigure[Overlap between the images]{\input{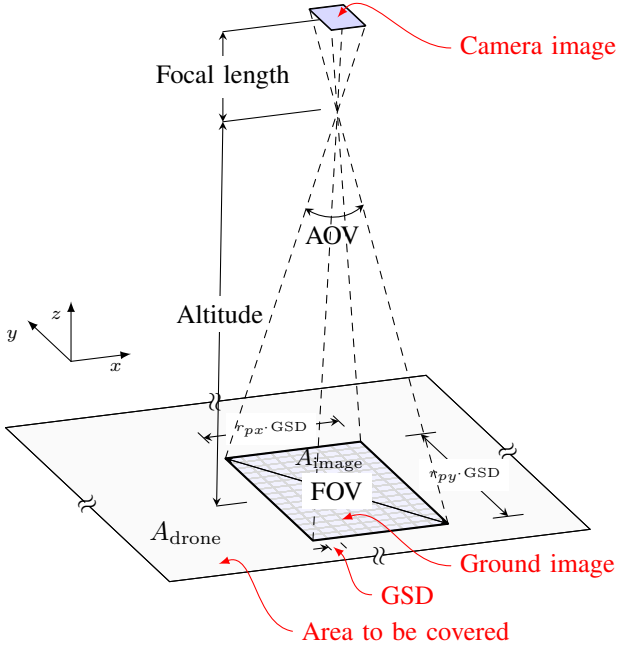}
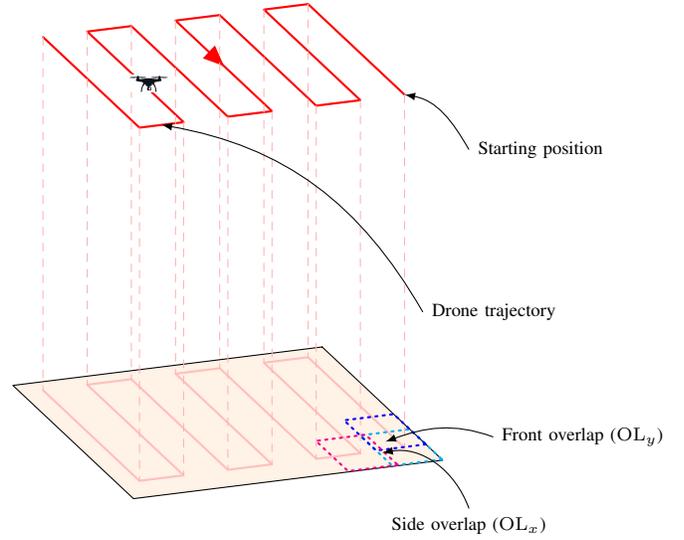\label{drone_image_overlap}}
\caption{illustration of image capture in drone surveillance scenario.} 
\end{figure*}

\item The throughput requirements depend on the type of mission, and, for image and video transmission applications, on the capability of the drone's on-board image processing module. In some cases, it is preferable to transmit raw (uncompressed) video or images. For example, when machine learning is performed in the cloud, large amounts of raw imagery needs be transmitted. 

\paragraph*{\textbf{Throughput requirement in video applications}}
Consider a scenario where it is required to scan a $3 \times  4$~km area with 20 drones. A 4K (4096 x 2160) resolution video with 
60 frames-per-second (FPS) requires a  throughput of 64 Mbps per drone, yielding a   sum throughput of 1.28 Gbps \cite{prabhu_TWC}.

\paragraph*{\textbf{Throughput requirement in image transmissions}}
Consider a scenario where a drone has to cover an area of $A_\mathrm{drone}$ [m$^2$] with a certain target spatial resolution (see Figure \ref{GSD_Focal_Length_Illustration}). 
The spatial resolution of the image depends on the ground sampling distance (GSD). 
The GSD is the distance between the centers of two neighboring pixels measured on the ground, as shown in Figure \ref{GSD_Focal_Length_Illustration}.  
The area covered by an image, say $A_\mathrm{image}$ [m$^2$], depends on the altitude and the properties of the camera,
specifically its aspect ratio, focal length, and angle of view (AOV). The 
flight duration, and the data rate required for transmitting an image with a certain target spatial resolution can be calculated as a function of altitude, GSD, and the
properties of the camera (focal length (FL) and pixel size (PS)). 
The altitude for a given GSD is given by
\begin{equation}\label{altitude_label}
H=\frac{\mathrm{GSD} \times \mathrm{FL}}{\mathrm{PS}}.
\end{equation}

Let $r_{px} \times r_{py}$ ($r_{px} < r_{py}$) be the camera resolution in pixels. 
The area covered by an image is $A_\mathrm{image}=r_{px}\cdot r_{py}\cdot \mathrm{GSD}^2$. 
Then, for a given AOV, the field of view (FOV) is calculated as
\begin{equation}\label{fov_label}
\mathrm{FOV} = 2 H \tan{\left(\frac{\mathrm{AOV}}{2}\right)} = \mathrm{GSD} \sqrt{r_{px}^2+r_{py}^2}.
\end{equation}
For example, assume that the target GSD of the mission is 2 cm/pixel, the camera's AOV is 60 degrees, and $\mathrm{PS}~=~2.3\times10^{-6}$~m. 
When $r_{px}\times r_{py}~=~$ 2664 $\times$ 1496, we have that the corresponding required $\mathrm{FOV}~=~$61~m, the required
focal length $\mathrm{FL}~=~2\times~10^{-3}$~m, the required drone altitude $H~=~$53~m, and the resulting area covered by an image $A_\mathrm{image}~=~$1594~m$^2$.
 
\begin{figure}[htbp]
\centering
\includegraphics[scale=.65]{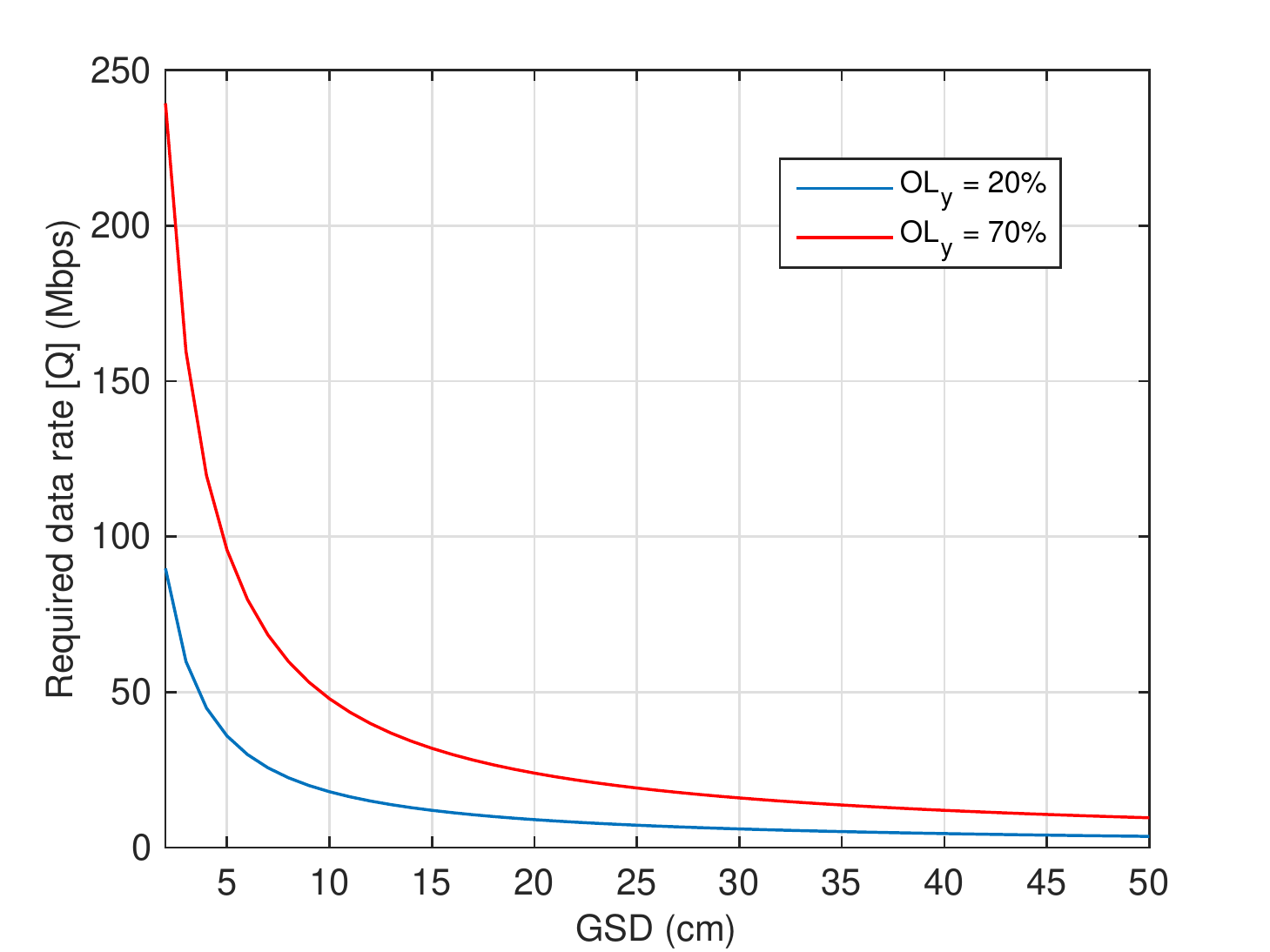}
\caption{Required data rate as a function of target GSD with $b~=~$24~bits/pixel, $\mathrm{CR}~=~$2:1, and $v~=~$20~m/s}
\label{data_rate_GSD}
\end{figure} 

Assume that the camera is oriented with the longest edge of the sensor 
parallel to the flight direction  (i.e. $y$-axis as shown in Figure \ref{drone_image_overlap}). 
Let $b$ be the number of bits per pixel and $\mathrm{CR}$ be the compression ratio. 
Let $\mathrm{OL}_y$ and $\mathrm{OL}_x$ be the
front and side image overlap (in percent), respectively, required by the
mission (see Figure \ref{drone_image_overlap}). 
Then the number of bits generated by an image is
\begin{equation}
 D_{\mathrm{image}}=\frac{r_{px}\cdot r_{py} \cdot b}{ \mathrm{CR}}
\end{equation}
and the time difference between two consecutive image captures is 
\begin{equation}
t=\frac{r_{py}\cdot \mathrm{GSD} \cdot (1-\mathrm{OL}_y)}{v},                                                          
\end{equation}
where $v$ is the drone speed.

For simplicity, let us assume that $\mathrm{OL}_x=$0. Then the instantaneous data rate required by the drone is  
\begin{equation}
Q=\frac{D_{\mathrm{image}}}{t}=\frac{r_{px} \cdot b \cdot v}{\mathrm{GSD} \cdot \mathrm{CR} \cdot (1-\mathrm{OL}_y)}.
\end{equation}

\begin{figure}[!htbp]
\centering
\includegraphics[trim = 2cm 4cm 2cm 2cm,clip,scale=.45]{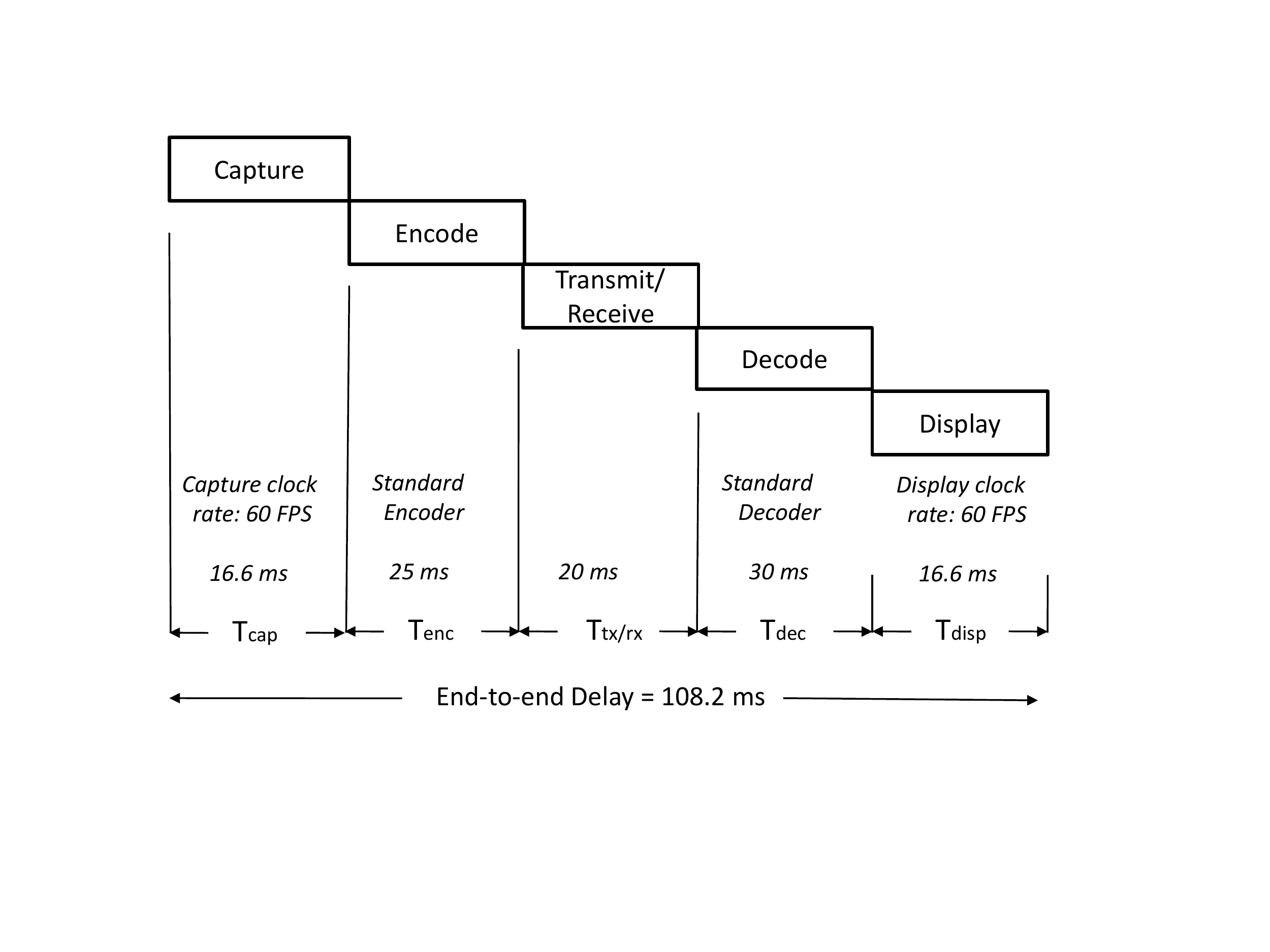}
\caption{End-to-end latency in video applications \cite{TI_Latency}.}
\label{Latency-Illustration}
\end{figure} 

Figure \ref{data_rate_GSD} shows the required data rate as  function of the target GSD with two different values of the image overlap. 
It can be seen that a higher spatial resolution requires a higher-throughput communication link.

\item Connectivity to multiple drones must be ensured
  simultaneously \cite{olsson2010,andre2014,OCHA,Northfield2013,gupta2016,Jaber_UAV,mademlis2019,Campion2018,diwu2017}.
  This requires either time- or frequency-sharing, which is inefficient. The spatial multiplexing capability of Massive MIMO can be exploited to simultaneously provide high throughput communication links to many drones.

\item Reliability and latency are important concerns in many
  applications. For example, in drone racing, the end-to-end delay should stay below a few tens of milliseconds \cite{Schneider2017}. 
  If a drone moves at 30 m/s, with a latency of 50 ms, when the pilot sees the video on the display the drone has moved 1.5 meters. Thus, the latency has to be kept as low as possible for drone control and navigation \cite{Kakar2017,TI_Latency}. Figure \ref{Latency-Illustration} illustrates the end-to-end latency in video applications. The main components that contribute to the increased latency are video encoding and decoding modules (greater than 50 ms processing time). By transmitting raw video (without any compression) via a high-throughput Massive MIMO based wireless link, it is possible to achieve low-latency in video applications.

\item High-mobility support is important \cite{asadpour2014} as typical flying speeds of drones range from 5 m/s (agricultural crop monitoring, site inspection) to 30 m/s (disaster management).

\end{itemize}

\subsection{ \textbf{Inadequacy of existing technologies}}

Existing wireless technologies such as WiFi and XBee-PRO are unsuitable for drone networks that involve high mobility and a large number of drones. Since these technologies were originally designed for indoor wireless access with very low mobility, they perform poorly in high-mobility conditions \cite{yanmaz2013}. Particularly, due to limitations in the PHY- and MAC-layer protocols, the latency is typically hundreds of milliseconds. Experimental results show that under LoS conditions, the maximum range of WiFi and XBee-PRO is 300--500 m (with 10 Mbps, single drone) and 1 km (with 250 kbps), respectively \cite{andre2014,hayat2016a}. 
Furthermore, more seriously, when multiple drones share  resources, due to the limitations of the MAC protocols, the throughput achieved per drone will decrease proportionally \cite{haitao2002}.

Recently there has been interest in utilizing 4G/5G cellular networks such as LTE for aerial communication \cite{yaliniz2016_1,guangyang2018,bergh2016,nguyen2018,bulut2018,kovacs2017,amorim2017_wcl,geraci2018_access,yzeng2018}. 
But cellular networks are not suitable for high throughput, long-range aerial communication purposes for the following reasons. Since the BS antennas typically tilted towards ground, measurement results show that, coverage is limited to low altitudes (below 100 m) and low elevation angles \cite{amorim2017_wcl,geraci2018,guangyang2018,yzeng2018,galkin2017}. 
Furthermore,   co-channel interference originating from neighboring cells will be a major problem \cite{amorim2018_wcl,kovacs2017}. 
Since existing cellular BSs are connected to the power grid, they may be unavailable in relevant emergency situations such as earth-quakes or massive flooding. 
Furthermore, cellular networks are not available in many relevant mountainous and sea environments. 
In addition, the mobility pattern of drones is different from that of terminals in cellular communications. Pitch, roll and yaw angles of the drones can change rapidly, and may result in poor connectivity due to polarization mismatch. 
 
A new, dedicated technology is required to provide connectivity to drones. 
A major advantage of developing a new technology for drone networks is that one can design the PHY- and MAC- layer protocols from a clean slate, accounting for the specific requirements of aerial communications.
\begin{figure*}[!t]
\centering
\begin{tikzpicture}[scale=.4]
\linespread{1};
\draw[thick,fill=gray!20] (1,-2) rectangle (9,4) node[pos=.5,text width=3cm] {\textbf{Control Information:} \\ Scheduling, Camera control, Trajectory control};

\draw[thick,fill=blue!20] (9,-2) rectangle (15,4) node[pos=.5,text width=2.15cm] {\textbf{Uplink}  \textbf{Pilot:} $K$ orthogonal sequences of length $\ge K$ symbols};

\draw[thick,fill=blue!30] (15,-2) rectangle (25,4) node[pos=.5,text width=2cm] {\textbf{Uplink Data:} Video, Image};

\draw[thick,fill=green!30] (25,-2) rectangle (31,4) node[pos=.5,text width=2.15cm] {\textbf{Downlink} \textbf{Pilot:} \\  $\ge$ 1 symbol (for uplink power control)};

\draw[thick,fill=green!30] (31,-2) rectangle (38,4) node[pos=.5,text width=2.5cm] {\textbf{Downlink Data:}\ \ \  e.g. Audio in case of disaster management };

\draw[decoration={brace,mirror,amplitude=10pt},decorate] (1,-3) -- node[below=12pt] {$\tau$ symbols} (37,-3);

\draw[decoration={brace,amplitude=10pt},decorate,yshift=8pt] (1,4) -- node[above=12pt] {$\tau_{\mathrm{ctrl}}$ } (9,4);

\draw[decoration={brace,amplitude=10pt},decorate,yshift=8pt] (9,4) -- node[above=12pt] {$\tau_{\mathrm{ul,p}}$} (15,4);

\draw[decoration={brace,amplitude=10pt},decorate,yshift=8pt] (15,4) -- node[above=12pt] {$\tau_{\mathrm{ul,d}}$ } (25,4);

\draw[decoration={brace,amplitude=10pt},decorate,yshift=8pt] (25,4) -- node[above=12pt] {$\tau_{\mathrm{dl,p}}$} (31,4);

\draw[decoration={brace,amplitude=10pt},decorate,yshift=8pt] (31,4) -- node[above=12pt] {$\tau_{\mathrm{dl,d}}$} (38,4);

\end{tikzpicture}
\caption{Allocation of symbols in TDD frame structure}
\label{frame_structure}
\end{figure*}
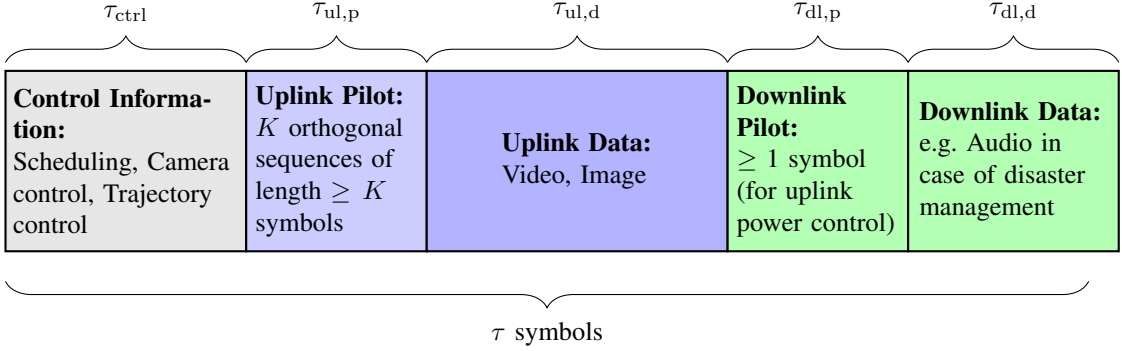

\section{ \textbf{Massive MIMO is a suitable technology for drone communications}}

 Massive MIMO is a multi-antenna multi-user wireless technology originally developed for cellular communications \cite{Marzetta16Book}. 
 The three main features of Massive MIMO are (i) array gain, which translates into a coverage extension; (ii) spatial multiplexing, which permits the service of many tens of terminals in the same time-frequency resource; and (iii) the handling of high mobility through exploitation of channel reciprocity and time-division duplex (TDD) operation \cite{Marzetta16Book,vieira2017}.  
 Substantially all signal processing complexity resides at the BS, rendering the terminals low-complexity. Massive 
 MIMO works well both in rich scattering and LoS environments \cite{Marzetta16Book,prabhu_TWC,harris2017}. 
 These features naturally make the technology suitable for drone communications.

While pilot contamination is known to be a limitation in multi-cell massive MIMO systems, for drone communications, due to high coherence bandwidth (assuming the antenna array is directed upwards into the sky) the coherence interval in samples is long and mutually orthogonal pilots to all drones can be afforded. Hence, pilot contamination is not a significant issue, particularly in scenarios where the drone density is low.

Field trials of Massive MIMO in high mobility have been performed for example in the pan-European FP7-MAMMOET project, and efficient hardware implementations have been demonstrated \cite{harris2015_bristol,harris2017}. In terms of digital circuit implementations, zero-forcing precoding and decoding of 8 terminals with 128 BS antennas over a 20 MHz bandwidth can be performed in real time at a power consumption of about 50 milliWatt \cite{HPrabhu2017}. Therefore, Massive MIMO GSs for drone communications can be realized at low cost and built from technology that is maturing \cite{perre2018}.

In hard-to-reach areas (for example, in mountainous and sea environments) and during natural disaster situations, Massive MIMO-based high-throughput networks can be deployed quickly to cover a large geographical area. 
A master aircraft equipped with an antenna array can act as a control station to gather information from a swarm of light-weight drones equipped with a single antenna. In urban environments, for providing coverage for high-altitude drones, the antenna array can be placed on top of high-rise buildings. 
Otherwise, the antenna array can be placed in existing cellular towers, but with appropriate tilting towards the sky.

\section{\textbf{What and how much Massive MIMO can provide for drone communications?}\hfill}
The fundamental principle of Massive MIMO operation is to obtain channel state information (CSI) between the antenna array and all terminals, and then apply appropriate signal processing algorithms. 
During the uplink training phase, all terminals simultaneously transmit predefined orthogonal pilot sequences to the GS. 
Upon receiving these pilots, the GS estimates the channels between the GS array and the terminals. 
TDD operation is used, which permits the exploitation of uplink-downlink channel reciprocity, 
as long as the uplink and downlink transmissions take place within the channel coherence time.

If the maximum drone speed is $v$ m/s, then the corresponding
coherence time is 
\begin{equation}
T_c\approx \frac{c}{2\cdot v\cdot f_c} \ \ \ \text{(s)},
\end{equation}
 where
$f_c$ is the carrier frequency and $c$ is the speed of light. With a  
coherence bandwidth of $B_c$ (Hz), the number of samples per
coherence interval is 
\begin{equation}
\tau = B_c T_c.
\end{equation}
For example, a coherence bandwidth of 3 MHz, a drone speed of 30
m/s, and a carrier frequency of 2.4 GHz, yield $\tau~=$~6250 samples.  

%
%
%
%

\begin{figure}
\centering
\tdplotsetmaincoords{60}{120}
\pgfmathsetmacro{\rvec}{2.36}
\pgfmathsetmacro{\thetavec}{65}
\pgfmathsetmacro{\phivec}{120}

\begin{tikzpicture}[scale=2,tdplot_main_coords]

\coordinate (O) at (0,0,0);

\tdplotsetcoord{P1}{\rvec}{\thetavec}{\phivec}
\tdplotsetcoord{P2}{2.6}{76}{125}

\draw[thick,->] (0,0,0) -- (2,0,0) node[anchor=north east]{\scriptsize$x$};
\draw[thick,->] (0,0,0) -- (0,.9,0) node[anchor=north west]{\scriptsize$y$};
\draw[thick,->] (0,0,0) -- (0,0,.9) node[anchor=south]{\scriptsize$z$};


	  \draw[latex-latex] (O) -- (P1)  node [midway,fill=white] {\footnotesize $d_k$};
	  \draw[latex-latex] (O) -- (P2)  node [midway,fill=white] {\footnotesize $d_j$};

\draw[-,dashed] (0,0,0) -- (-.48,.12,-0.20);
\draw[-,dashed] (-.48,.12,-0.2) -- (-.48,.12,.01);

\draw[-,dashed] (0,0,0) -- (-.48,.5,-0.20);
\draw[-,dashed] (-.48,.5,-0.2) -- (-.48,.5,.07);

\draw[-stealth,color=black] (.585,-.51,0) -- (0,-.51,0);
\draw[-stealth,color=black] (.8,-.51,0) -- (1.6,-.51,0);

\draw[-,color=black] (0,-.45,0) -- (0,-.575,0);
\draw[-,color=black] (1.61,-.45,0) -- (1.61,-.575,0);

\draw[-,color=black] (0,-.25,0) -- (0,-.35,0);
\draw[-,color=black] (.2,-.25,0) -- (.2,-.35,0);

\draw[-stealth,color=black] (.35,-.29,0) -- (.2,-.29,0);
\draw[-stealth,color=black] (-.1,-.29,0) -- (0,-.29,0);

\node at (.1,-.3,0) {\tiny $\delta$};
\node at (.7,-.51,0) {\scriptsize $(M-1)\delta$};

\tdplotdrawarc[->]{(O)}{0.17}{0}{\phivec}{anchor=north}{\scriptsize$\phi_{k}$}

\tdplotsetthetaplanecoords{\phivec}
\tdplotdrawarc[->,tdplot_rotated_coords]{(0,0,0)}{0.2}{0}{\thetavec}{anchor=south west}{\scriptsize$\theta_{k}$}

\tdplotdrawarc[->]{(O)}{0.6}{0}{106}{anchor=north}{\scriptsize$\phi_{j}$}

\tdplotsetthetaplanecoords{125}
\tdplotdrawarc[->,tdplot_rotated_coords]{(0,0,0)}{0.6}{0}{76}{anchor=south west}{\scriptsize$\theta_{j}$}


\draw[] (0,0,0) node[circle,fill,inner sep=1.25pt,label=above:](a){} -- (.2,0,0);

\draw[] (0,0,-0.125) -- (0,0,0.125);

\draw[] (.2,0,0) node[circle,fill,inner sep=1.25pt,label=above:](a){} -- (.2,0,0);

\draw[] (.2,0,-0.125) -- (.2,0,0.125);

\draw[] (.4,0,0) node[circle,fill,inner sep=1.25pt,label=above:](a){} -- (.2,0,0);

\draw[] (.4,0,-0.125) -- (.4,0,0.125);

\draw[] (.6,0,0) node[circle,fill,inner sep=1.25pt,label=above:](a){} -- (.2,0,0);

\draw[] (.6,0,-0.125) -- (.6,0,0.125);

\draw[] (0.8,0,0) node[circle,fill,inner sep=1.25pt,label=above:](a){} -- (.2,0,0);

\draw[] (.8,0,-0.125) -- (.8,0,0.125);

\draw[] (1.0,0,0) node[circle,fill,inner sep=1.25pt,label=above:](a){} -- (.2,0,0);

\draw[] (1,0,-0.125) -- (1,0,0.125);

\draw[] (1.2,0,0) node[circle,fill,inner sep=1.25pt,label=above:](a){} -- (.2,0,0);

\draw[] (1.2,0,-0.125) -- (1.2,0,0.125);

\draw[] (1.4,0,0) node[circle,fill,inner sep=1.25pt,label=above:](a){} -- (.2,0,0);

\draw[] (1.4,0,-0.125) -- (1.4,0,0.125);

\draw[] (1.6,0,0) node[circle,fill,inner sep=1.25pt,label=above:](a){} -- (.2,0,0);

\draw[] (1.6,0,-0.125) -- (1.6,0,0.125);

\begin{scope}[xshift=0,yshift=55]

\node at (P1) {\includegraphics[scale=.05]{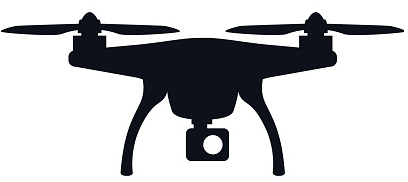}};

\node at (P2) {\includegraphics[scale=.05]{drone_icon}};

\end{scope}

\end{tikzpicture}\caption{Canonical  geometric model. The elevation and azimuth angles determine the channel responses of the drones}
\label{geometric_model}
\end{figure}

\textbf{TDD frame structure:} Figure \ref{frame_structure} illustrates the TDD-based frame structure where $\tau_{\mathrm{ctrl}}$, $\tau_{\mathrm{ul,p}}$, $\tau_{\mathrm{ul,d}}$, $\tau_{\mathrm{dl,p}}$, and $\tau_{\mathrm{dl,d}}$ denote the number of symbols used for control channel information, uplink pilot transmission, uplink data transmission, downlink pilot transmission, and   downlink data transmission, respectively, and $\tau = \tau_{\mathrm{ctrl}}+\tau_{\mathrm{dl,p}}+\tau_{\mathrm{dl,d}}+\tau_{\mathrm{ul,p}}+\tau_{\mathrm{ul,d}}$. 

\textbf{Control information:} The command and control information from the GS is transmitted through downlink control channels. 
The control information consists of important and critical data such as synchronization, scheduling information, and direction or trajectory control \cite{nasa_report2012,she2019}. With Massive MIMO, further functionalities can be added, for example: control signals for re-orienting drone's antenna towards the GS and selection of antenna polarization. In order to achieve high reliability, instead of spatial multiplexing, it is preferred to transmit the control information on orthogonal resource elements \cite{she2019}. 

\textbf{Uplink pilot transmission:} If there are $K$ drones in the system, then the number of symbols used for uplink pilot transmission should be greater than or equal to $K$ \cite{Marzetta16Book}. The pilot power of the drones is chosen based on the maximum possible distance and the worst-case combined effect of antenna gain and polarization mismatch factor (averaged over all antenna elements), so that a pre-determined received pilot signal-to-noise ratio (SNR) is maintained at the GS. Polarization mismatch losses, per antenna, can be severe in some cases. However, the overall loss can be made small by an appropriate choice of polarization and orientation of the GS array elements.

\textbf{Uplink data transmission:} During uplink data transmission, all drones apply channel inversion power control in order to maintain a pre-determined target data SNR at the GS (averaged over all GS antenna elements). 

\textbf{Downlink pilot transmission:} Through broadcasting of (non-directional) pilots by the GS, the drones can measure their path loss. 
Note that one downlink pilot is sufficient  in the downlink training phase.

\begin{figure}[!htbp]
\centering
\includegraphics[scale=.65]{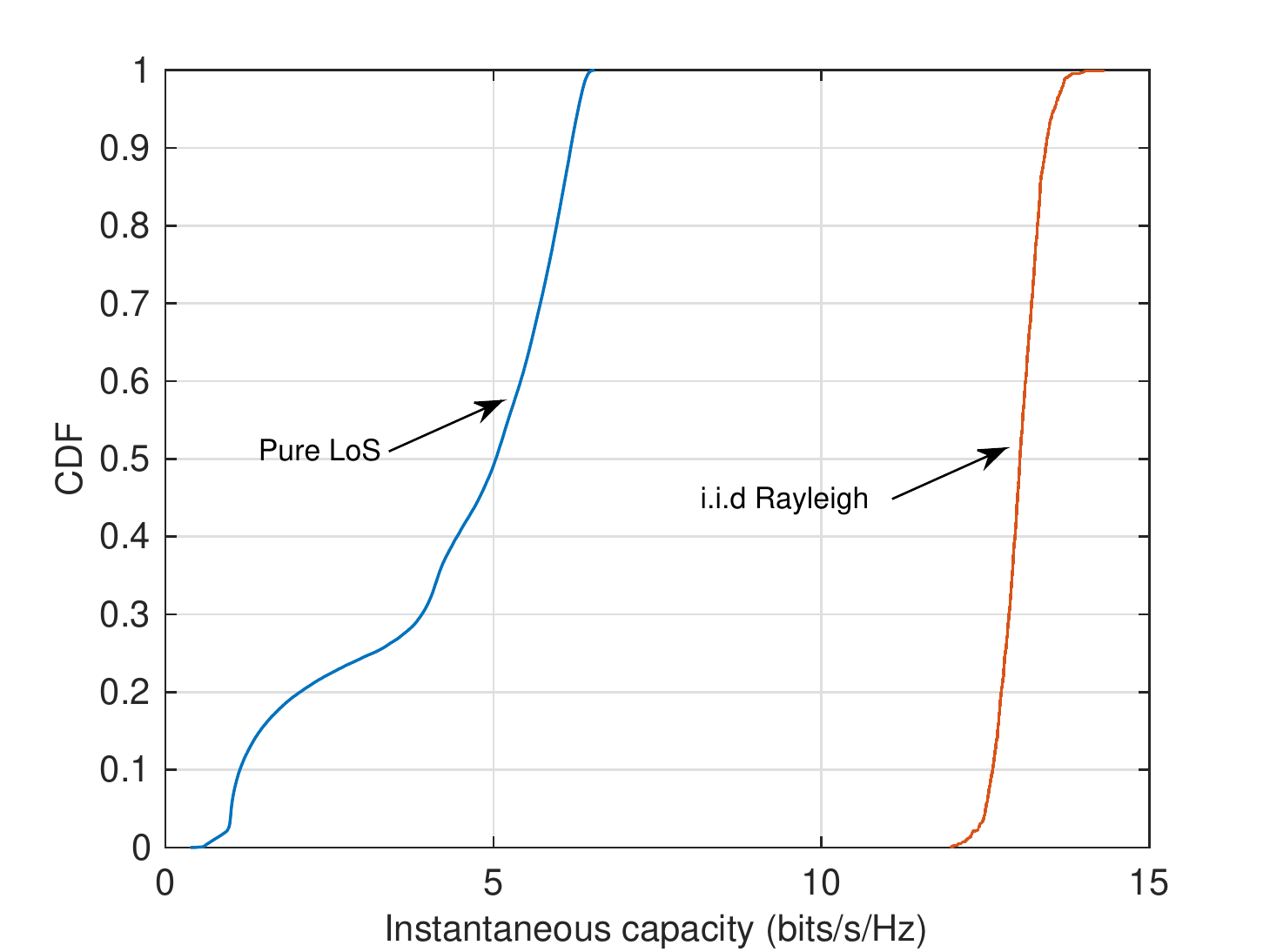}
\caption{CDF of instantaneous uplink capacity with pure LoS and i.i.d.\ Rayleigh fading channels with $M=$100 antennas and $K=$20 terminals at 2.4 GHz carrier frequency. The terminal positions are assumed to be uniformly and independently distributed within a sphere of radius 500 m. Uplink channel inversion power control is applied to achieve equal SNR (0 dB) for all terminals.}
\label{Mag_LoS_Rayleigh_comparison}
\end{figure} 

For the examples presented next, we assume that the GS comprises a uniform linear or rectangular array with $M$ antennas. The drones, in contrast, are equipped with a single antenna. The GS uses fully digital maximum-ratio combining (MRC) processing to decode the data transmitted by the drones.

For  given channel vectors of $K$ drones, the instantaneous uplink capacity achieved by the $k$-th drone is given by
\begin{equation}
 S_k = \log_2\left(1+\frac{p_{uk}\norm{\boldsymbol{g}_{k}}^4}{\sum_{j=1,j\neq k}^K p_{uj}|\boldsymbol{g}_{k}^H \boldsymbol{g}_j|^2+\norm{\boldsymbol{g}_k}^2}\right),
\end{equation}
where the $k$-th drone's channel vector $\boldsymbol{g}_k = [g_{k1}\ g_{k2}\ ...\ g_{kM}]^T$ and $p_{uj}$ is the transmit power (normalized by the receiver noise power) of $j$-th drone.
We consider that there is LoS propagation (no multipath) between the GS antenna array and the drones. The channel between the $l$-th GS antenna and the $k$-th drone's antenna is given by
\begin{equation}\label{channel_los}
g_{kl} = \sqrt{\beta_{k}\chi_{kl}}\exp{\left(-\frac{i 2\pi \left(d_k+(l-1)\delta\sin \theta_k \cos \phi_k\right)}{\lambda}\right)},
\end{equation}
where $\beta_k$ represents free-space path loss, $\chi_{kl}$ represents polarization mismatch loss and antenna gain, $d_k$ is
the distance between the first GS antenna and the drone (see Figure \ref{geometric_model}), $\delta$ is the antenna 
element spacing, and $\lambda$ is the operating wavelength. 

In practice, the Massive MIMO system performance will lie between pure LoS and rich scattering (i.e., i.i.d.\ Rayleigh fading) environments. In rich scattering, which is typical of a cellular environment, the channel coefficients are given by
\begin{equation}\label{channel_ray}
g_{kl} \sim \mathcal{CN}(0, \beta_k), \ \ l=1,2,... M.
\end{equation}

Favorable channel conditions hold with high probability in  rich scattering, where  channels become asymptotically orthogonal as the number of antennas grows. Interestingly, favorable channel condition can   be achieved with large numbers of antennas even in pure LoS scenarios as well \cite[Chap.~7]{Marzetta16Book}. 
Figure \ref{Mag_LoS_Rayleigh_comparison} shows the CDF of the instantaneous capacity achieved by the terminals with two different channel models as given in \eqref{channel_los} and \eqref{channel_ray}. 
With pure LoS, since the channel vectors of different terminals are highly correlated, the capacity of the LoS channel is significantly lower than that of the i.i.d. Rayleigh fading channel. The practical air-to-ground channels are characterized by a small number of multi-path components which is close to the pure LoS channel \cite{newhall2003,rajashekar2018}. 
Thus, the LoS assumption considered in this paper provides worst-case system design for Massive MIMO based drone communication system.

\begin{figure}[!htbp]
\centering
\includegraphics[scale=.65]{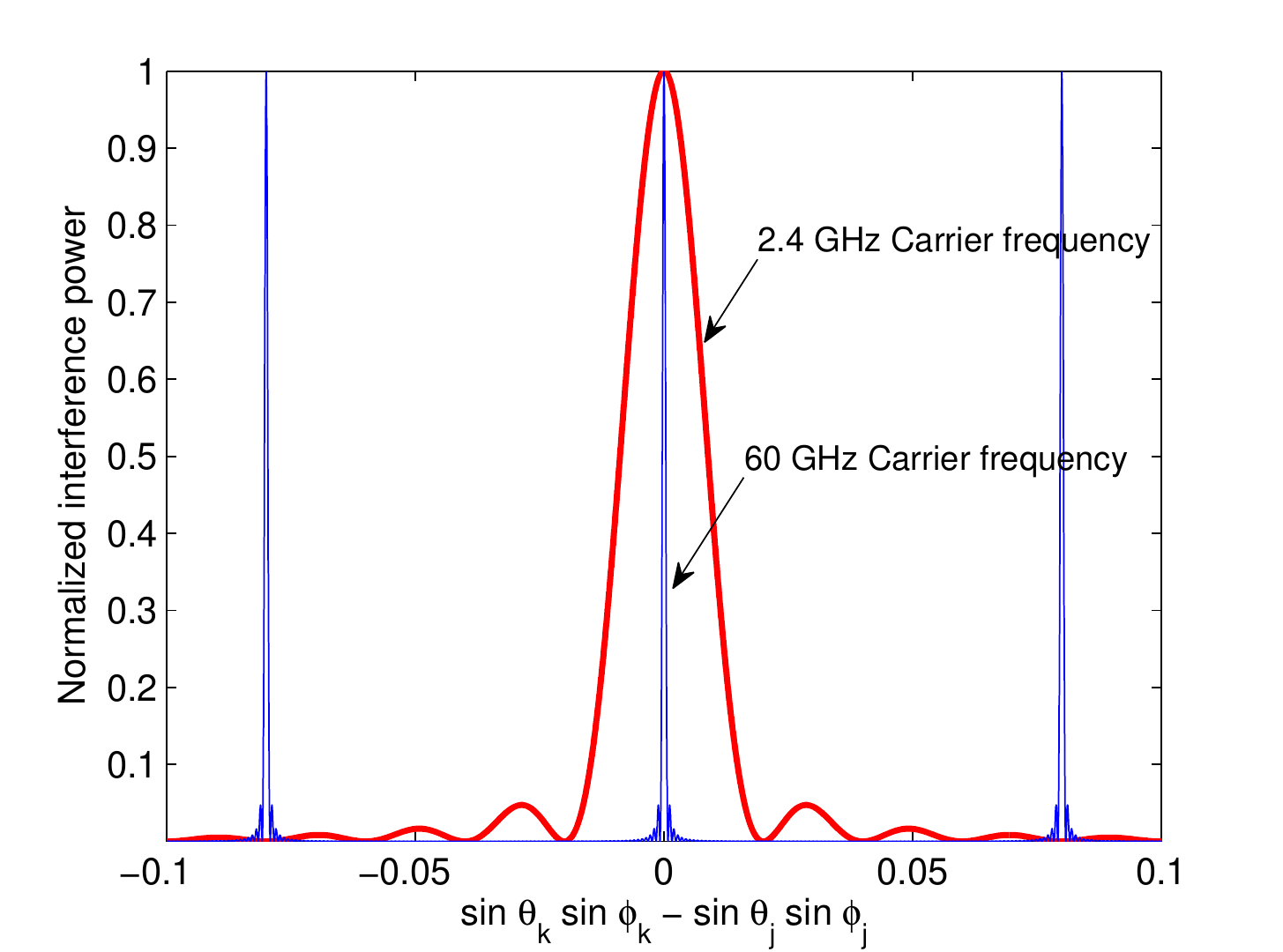}
\caption{Interference as function of angular separation, quantified in terms of $\sin \theta_k \sin \phi_k - \sin \theta_j \sin \phi_j$ with 100 GS antennas. 
The aperture length is 6.25 m, the same for both carriers (2.4 and 60 GHz). The spacing between the elements is half a wavelength for 2.4 GHz and 12.5 wavelengths at 60 GHz. 
At 60 GHz, the mainlobe is 25 times narrower than at 2.4 GHz. 
Note that the actual range of this quantity is between -2 to 2 but for clarity, only the range  [-0.1,0.1] is shown.}
\label{Mag_lobe_V3}
\end{figure}

\begin{figure*}[!htbp]
\centering
\subfigure[
Sum throughput vs. number of drones. The drones are uniformly and randomly located in a spherical shell with inner radius 20 m and outer radius 500 m. For tractability reasons we assume that the GS array elements are identically oriented. The graph is generated using Eqn. \eqref{asym_high_snr_mrc}. Due to the channel inversion power control, the same data SNR is maintained at each receive antenna. The distribution of the corresponding uplink transmit power is shown in Figure~\ref{Magazine_CDF_power}.]{\includegraphics[scale=.585]{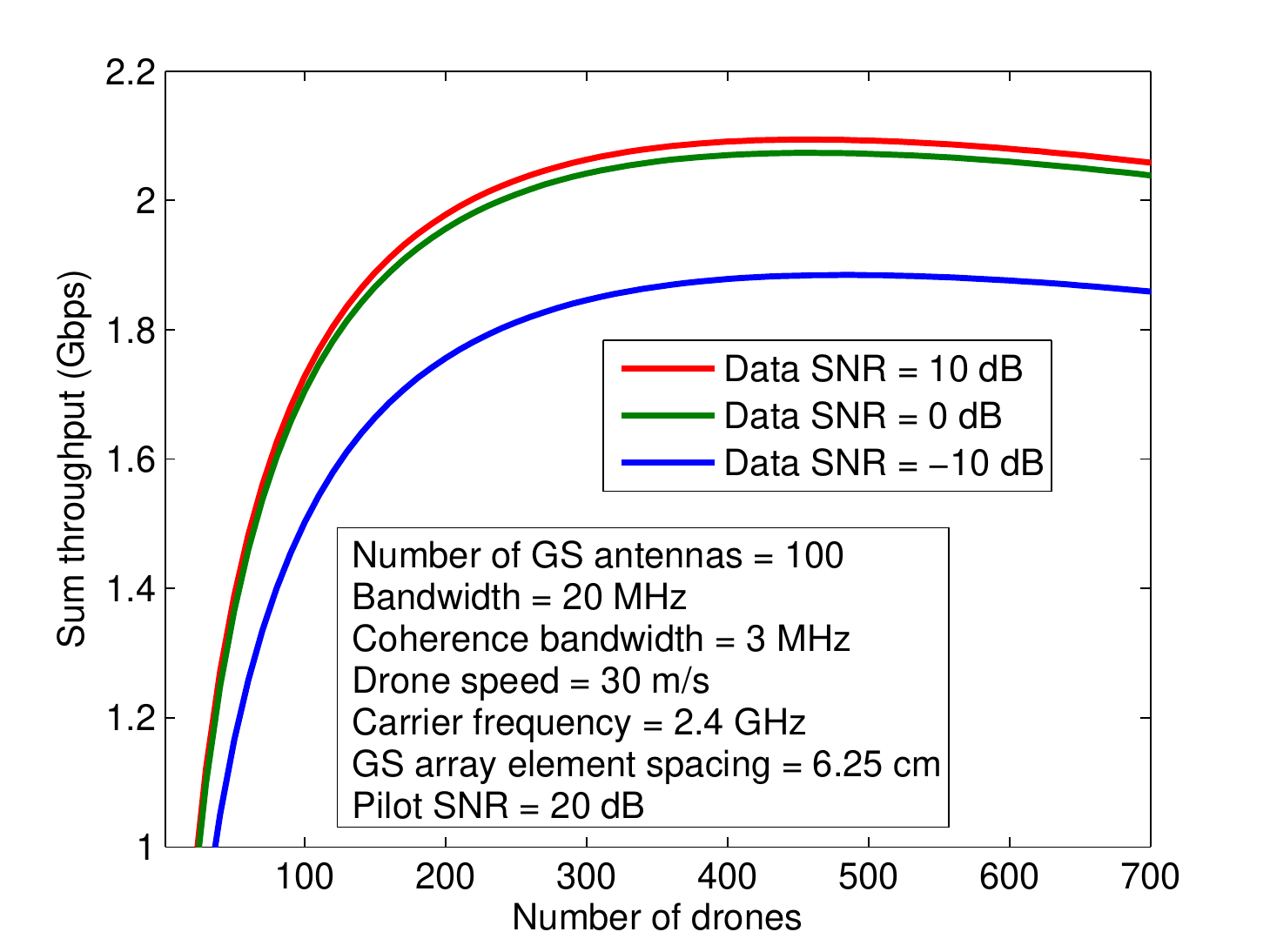}\label{Mag_sumThroughput_K}}\vspace{-.135cm}
\subfigure[Cumulative distribution function of the uplink transmit power required to maintain 0 dB data SNR at the receiver with different orientations of the GS antennas  (these are cross-dipole antennas). The randomness comes from randomness in the location and the orientation of the drones. The drone's roll, pitch, and yaw angles are independently and uniformly distributed. Power control is applied to maintain a pre-determined target data SNR (averaged over all GS elements, see \cite{prabhu_TWC}). In case of pseudo-randomly oriented GS antennas, the SNR will be different at different elements in the array. A drone is within coverage of the GS   if the drone's uplink power is sufficient to   compensate for the distance-dependent path-loss and polarization
  mismatch loss.]{\includegraphics[scale=.585]{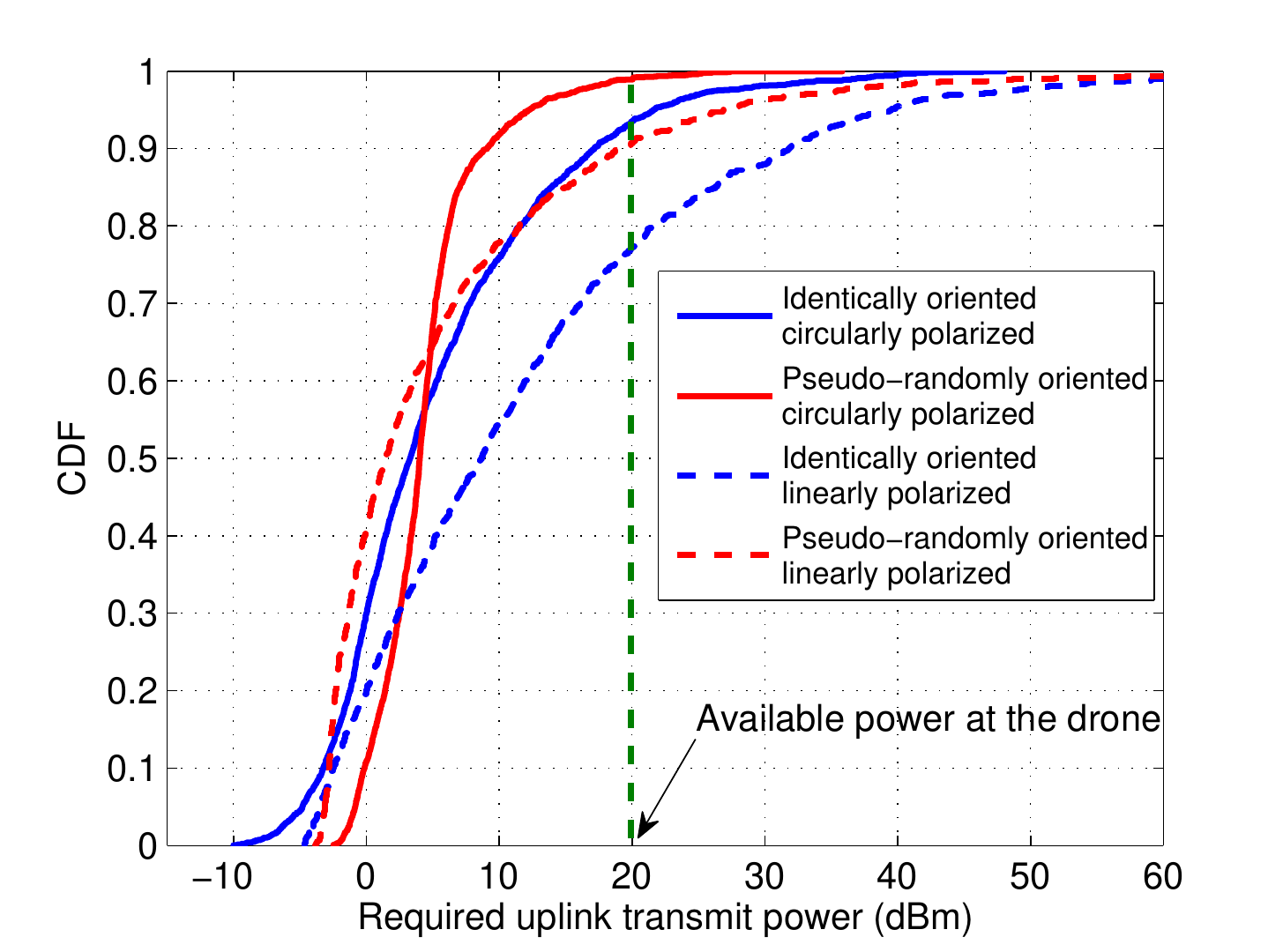}\label{Magazine_CDF_power}}
\caption{Examples of throughput and coverage performance.} 
\end{figure*} 

We use the ergodic rate as performance metric, that is, averaging the instantaneous throughput over all possible drone positions. The justification is that first, owing to the channel-inversion power control, the instantaneous throughput does not depend much on the range, and, second, that the multi-user interference experienced by a drone depends on the relative positions of the other drones and the strength of this interference fluctuates quickly as the drones move. For example, consider the geometric model    in Figure \ref{geometric_model}. The normalized interference power caused by the $j$-th drone towards the $k$-th drone can be expressed as \cite{prabhu_TWC}
\begin{align}
\frac{|\boldsymbol{g}_{k}^H\boldsymbol{g}_j|^2}{\beta_k\beta_j \chi_{k}\chi_{j}}= 
&\frac{\mathrm{sinc}^2\left(M\frac{\delta}{\lambda} 
(\sin{\theta_k}\cos{\phi_k}-\sin{\theta_j}\cos{\phi_j})\right)}
{\mathrm{sinc}^2\left(\frac{\delta}{\lambda} 
(\sin{\theta_k}\cos{\phi_k}-\sin{\theta_j}\cos{\phi_j})\right)},
\end{align}
where $\mathrm{sinc}(x)=\frac{\sin(\pi x)}{\pi x}$. As an illustration, for the geometric model in
 Figure~\ref{geometric_model}, Figure~\ref{Mag_lobe_V3} shows the strength of the interference that one drone is causing to another, as a function of their angular separation, i.e. $\sin \theta_k \sin \phi_k - \sin \theta_j \sin \phi_j$. 
 
\subsection*{\hfil \textbf{Simultaneous high-throughput transmission from a large number of drones }\hfil}
For analytical tractability reasons, assuming that the GS array elements are identically oriented (for the general case, see the next section), for a given distributions of the drones, a lower bound on the uplink ergodic rate (in bits/sec) achieved per drone is given by \cite{prabhu_TWC},
\small\begin{align}\label{asym_high_snr_mrc}
S = B &\left(1-\frac{2\cdot v\cdot f_c\left(K+\tau_{\mathrm{dl}}+\tau_{\mathrm{ctrl}}\right)}{B_c\cdot c} \right)\nonumber\\
&\times \log_2\left(1+\frac{M\rho_u}{\rho_u(K-1)\left(1+\frac{\Omega}{M}\right)+1+\frac{\kappa \ \chi_{wc}}{\rho_u\rho_p}(1+K\rho_u)}\right),
\end{align}\normalsize
where $B$ is the bandwidth and $\tau_\mathrm{dl}=\tau_\mathrm{dl,p}+\tau_\mathrm{dl,d}$ is the number of symbols used for downlink transmission. $\chi_{wc}$ is the lowest possible
value of the combined effect of antenna gain and polarization mismatch and $\kappa$ is the average antenna
gain and polarization mismatch; $\kappa \ \chi_{wc} < 1$.

The quantity $\Omega$ can be interpreted as the correlation between the spatial signatures of two different drones. Consider that the drone positions are distributed uniformly at random within
  a spherical shell with inner radius $R_{\mathrm{min}}$ and outer
  radius $R$, where the inner radius is much greater than the array
  aperture. When the inner radius of the sphere is comparable to the outer radius, the quantity $\Omega$ is given by
\begin{equation}
\Omega = \sum_{l=1}^M\sum_{l'=1,l'\neq l}^M \mathrm{sinc}^2\left(2(l-l')\frac{\delta}{\lambda}\right).
\end{equation}
Note that with element spacing equal to an integer multiple of one-half wavelength, the quantity $\Omega$ becomes zero. For any other distributions of the drone positions the rate expression as given in \eqref{asym_high_snr_mrc} remains the same except that the quantity $\Omega$ which may be different. Therefore, the expression in \eqref{asym_high_snr_mrc} can be used to estimate the throughput performance of practical deployment scenarios.

By virtue of the spatial multiplexing capability of Massive MIMO, many
drones can simultaneously transmit high-resolution imagery to the
GS. Figure \ref{Mag_sumThroughput_K} exemplifies the sum throughput
for different  numbers of drones, using a 20 MHz bandwidth and 100 GS
antennas. The sum throughput increases
up to a certain number of drones and then decreases. This is because the pilot overhead per coherence block is proportional to the number of drones, such that the number of samples available for uplink data transmission decreases with
 the number of drones. Moreover, it can be observed that the sum throughput is almost the same when the
data SNR equals 0 dB and 10 dB. This is so because when increasing the SNR, the sum throughput saturates as the communication becomes interference limited.

%
%
%
%

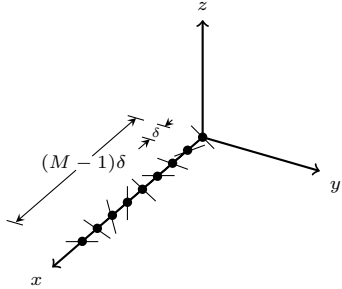
\begin{figure}
\centering
\tdplotsetmaincoords{60}{120}
\pgfmathsetmacro{\rvec}{2.36}
\pgfmathsetmacro{\thetavec}{65}
\pgfmathsetmacro{\phivec}{120}

\begin{tikzpicture}[scale=2,tdplot_main_coords]

\coordinate (O) at (0,0,0);

\tdplotsetcoord{P1}{\rvec}{\thetavec}{\phivec}
\tdplotsetcoord{P2}{2.6}{76}{125}

\draw[thick,->] (0,0,0) -- (2,0,0) node[anchor=north east]{\scriptsize$x$};
\draw[thick,->] (0,0,0) -- (0,.9,0) node[anchor=north west]{\scriptsize$y$};
\draw[thick,->] (0,0,0) -- (0,0,.9) node[anchor=south]{\scriptsize$z$};





\draw[-stealth,color=black] (.585,-.51,0) -- (0,-.51,0);
\draw[-stealth,color=black] (.8,-.51,0) -- (1.6,-.51,0);

\draw[-,color=black] (0,-.45,0) -- (0,-.575,0);
\draw[-,color=black] (1.61,-.45,0) -- (1.61,-.575,0);

\draw[-,color=black] (0,-.25,0) -- (0,-.35,0);
\draw[-,color=black] (.2,-.25,0) -- (.2,-.35,0);

\draw[-stealth,color=black] (.35,-.29,0) -- (.2,-.29,0);
\draw[-stealth,color=black] (-.1,-.29,0) -- (0,-.29,0);

\node at (.1,-.3,0) {\tiny $\delta$};
\node at (.7,-.51,0) {\scriptsize $(M-1)\delta$};






\draw[] (0,0,0) node[circle,fill,inner sep=1.25pt,label=above:](a){} -- (.2,0,0);

\draw[rotate=45] (0,0,-0.125) -- (0,0,0.125);

\draw[] (.2,0,0) node[circle,fill,inner sep=1.25pt,label=above:](a){} -- (.2,0,0);

\begin{scope}[shift={(.145,-.15)},rotate=25]
\draw[rotate=85] (.2,0,-0.125) -- (.2,0,0.125);
\end{scope}

\draw[] (.4,0,0) node[circle,fill,inner sep=1.25pt,label=above:](a){} -- (.2,0,0);

\begin{scope}[shift={(.025,-.325)},rotate=25]
\draw[rotate=45] (.4,0,-0.125) -- (.4,0,0.125);
\end{scope}

\draw[] (.6,0,0) node[circle,fill,inner sep=1.25pt,label=above:](a){} -- (.2,0,0);

\begin{scope}[shift={(.2,-.525)},rotate=45]
\draw[rotate=45] (.6,0,-0.125) -- (.6,0,0.125);
\end{scope}

\draw[] (0.8,0,0) node[circle,fill,inner sep=1.25pt,label=above:](a){} -- (.2,0,0);

\begin{scope}[shift={(1.73,.525)},rotate=182]
\draw[rotate=45] (.8,0,-0.125) -- (.8,0,0.125);
\end{scope}

\draw[] (1.0,0,0) node[circle,fill,inner sep=1.25pt,label=above:](a){} -- (.2,0,0);

\begin{scope}[shift={(.13,.075)},rotate=0]
\draw[rotate=0] (1,0,-0.05) -- (1,0,0.21);
\end{scope}

\draw[] (1.2,0,0) node[circle,fill,inner sep=1.25pt,label=above:](a){} -- (.2,0,0);

\begin{scope}[shift={(.012,-.15)},rotate=10]
\draw[rotate=5] (1.2,0,-0.02) -- (1.2,0,0.25);
\end{scope}

\draw[] (1.4,0,0) node[circle,fill,inner sep=1.25pt,label=above:](a){} -- (.2,0,0);

\begin{scope}[shift={(-.1375,-1)},rotate=10]
\draw[rotate=45] (1.4,0,-0.125) -- (1.4,0,0.125);
\end{scope}

\draw[] (1.6,0,0) node[circle,fill,inner sep=1.25pt,label=above:](a){} -- (.2,0,0);

\begin{scope}[shift={(0.565,-1.4)},rotate=90]
\draw[rotate=0] (1.6,0,-0.125) -- (1.6,0,0.125);
\end{scope}





\end{tikzpicture}\caption{Illustration of antenna array with pseudo-randomly oriented elements}
\label{geometric_model_random}
\end{figure}

 \begin{figure*}[!htbp]
\centering
\subfigure[With identically oriented array elements as shown in Figure \ref{geometric_model}.]{\includegraphics[scale=.585]{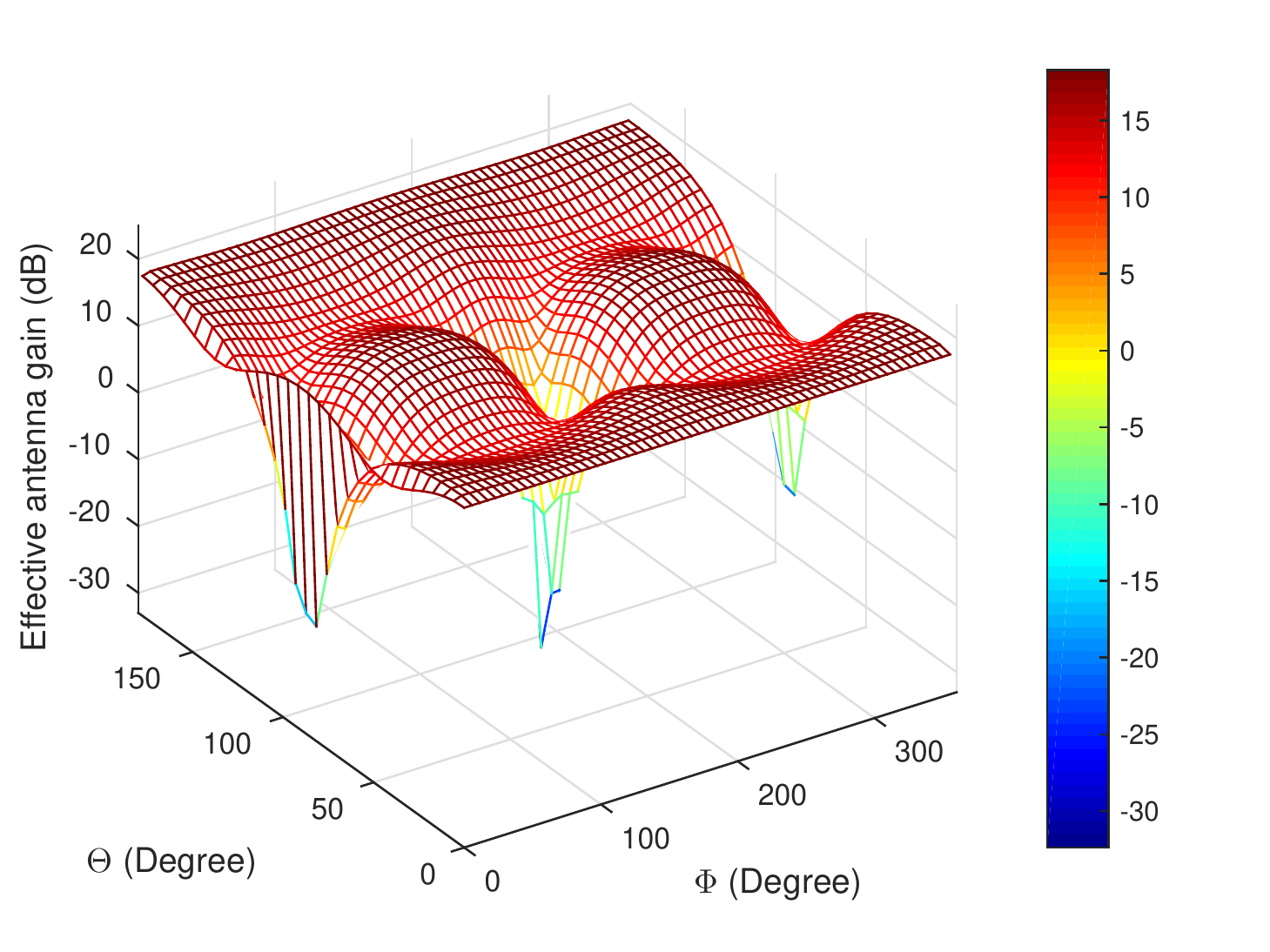}\label{Mag_Eff_gain_Cir_I}}
\subfigure[With pseudo-randomly oriented GS array elements as shown in Figure \ref{geometric_model_random}.]{\includegraphics[scale=.585]{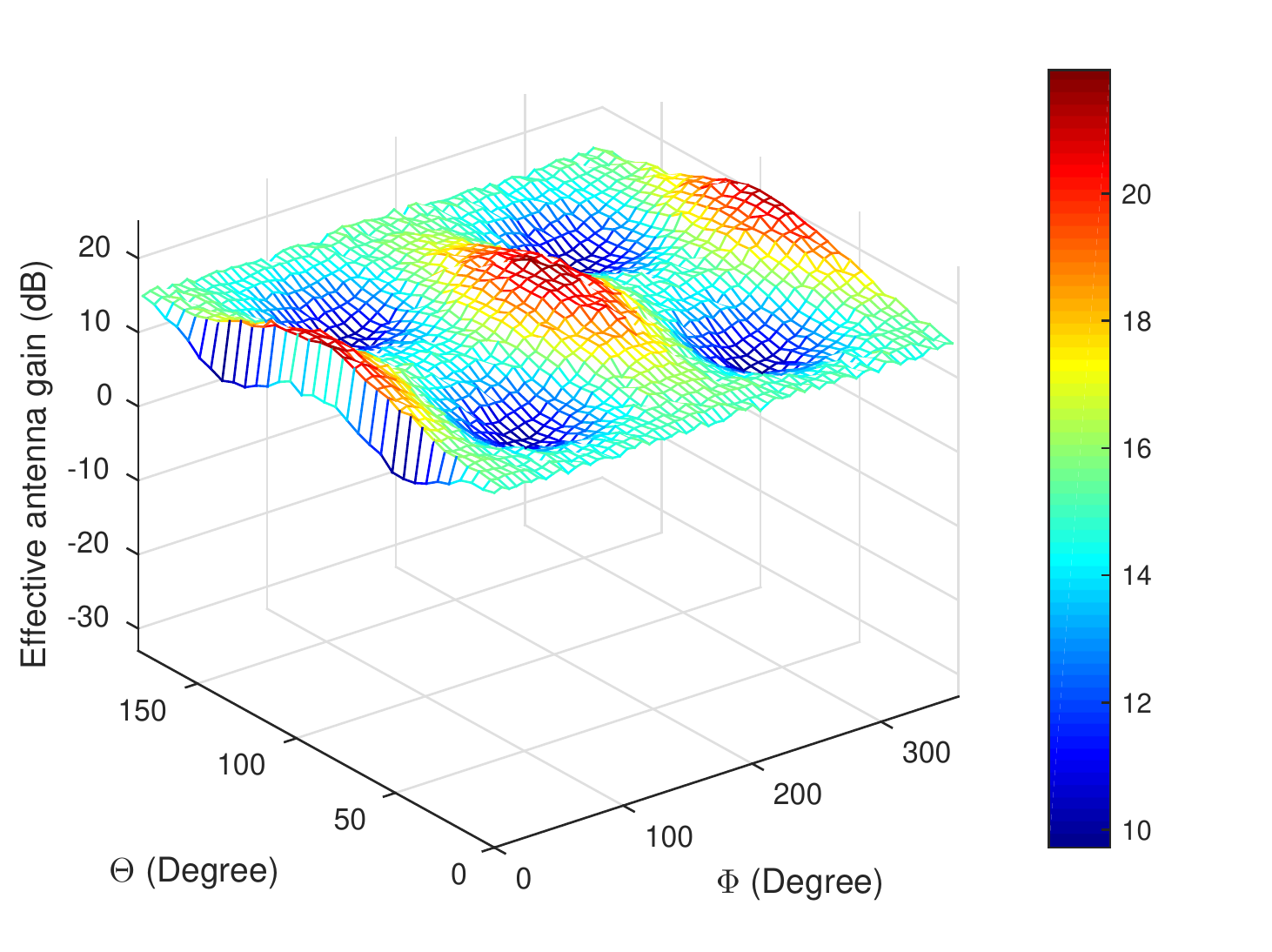}\label{Mag_Eff_gain_Cir_NI}}
\caption{Effective antenna gain with identically and pseudo-randomly oriented GS array elements. The effective antenna gain is 
$\chi_{k} = \sum_{l=1}^{M} \chi_{kl}$,
where $\chi_{kl}$ is the antenna gain between the $l$-th GS antenna and the $k$-th drone.} 
\end{figure*}

\subsection*{\hfil \textbf{Pseudo-randomly oriented circularly polarized antenna elements enable stable link conditions}\hfil}

Coverage in a 3D space requires careful antenna design \cite{yanmaz2013,asadpour2014}. The GS array may comprise simple antennas such as cross-dipoles. If all antenna elements are independently oriented in arbitrary directions, then the probability of having a high mismatch loss to all antenna elements can be greatly reduced. Figure \ref{Magazine_CDF_power} shows the cumulative probability distribution of the required transmit power, assuming a noise spectral density of -167 dBm/Hz, and a target SNR of 0 dB. With an available power budget of 100 mW (20 dBm), the probability of coverage is 78\% with identically oriented, linearly polarized antennas, but 99.99\% with circularly polarized, pseudo-randomly oriented antennas. 

The differences in coverage probability stem from signal losses due to polarization mismatches between the GS antennas and the antenna at the drone. By pseudo-randomly orienting the antennas in the GS array as shown in Figure \ref{geometric_model_random}, the coverage probability can be greatly improved irrespective of the location and orientation of the drones. To exemplify further, Figures \ref{Mag_Eff_gain_Cir_I} and \ref{Mag_Eff_gain_Cir_NI} show the effective antenna gain for varying azimuth and elevation angles with identically and pseudo-randomly oriented GS array elements, respectively. It can be observed that with identically oriented GS array elements, the effective antenna gain becomes very low (below $-30$ dB) at certain azimuth and elevation angles. Further, the antenna gain varies between -30 dB to 20 dB. On the other hand, with pseudo-randomly oriented GS array elements, the effective antenna gain stays within 10--20 dB.

The array geometry also determines the throughput performance, since it determines the angular resolvability. A rectangular array is preferred over linear and circular arrays due to its 3D resolution and reduced space occupancy. For 2.4 GHz and 60 GHz carrier frequencies, with half-wavelength spacing, the space occupied by a rectangular array with 400 elements is approximately 1.25 m x 1.25 m and 5 cm x 3 cm, respectively.

Moreover, in practice, the choice of element spacing is also important. It is observed that, for uniformly
distributed drone positions inside a spherical shell, and MRC signal
processing, the throughput is maximized when the GS array elements are
spaced at a distance equal to an integer multiple of one-half
wavelength \cite{prabhu_TWC}. For other distributions of drone
positions, the optimal antenna spacing is in general different.

\subsection*{\hfil \textbf{High mobility support}\hfil}

In many environments, the drone speed has only a slight impact on the
data rate. For example, in over-water and mountainous settings, with
 100 drones, at 2.4 GHz carrier frequency, 3 MHz coherence bandwidth, and
 if 90\% of samples per coherence interval are used for uplink transmission,  if
  we change the drone speed from
0 m/s to 30 m/s, from Eqn. \eqref{asym_high_snr_mrc} we can observe that the reduction in throughput is less than
2\%. In other environments and at larger carrier frequencies, this loss   becomes
more significant.

By exploiting the characteristics of LoS propagation, one might reduce
the channel estimation overhead. For example, if a drone is moving
along a known trajectory as shown in Figure \ref{drone_image_overlap}, the channel will experience a predictable
phase shift. Hence, if the channel is known at a specific point in
time, it may be predicted during many wavelengths of motion without
the transmission of additional pilots. This could further increase the
number of samples available for data transmission.

\subsection*{\hfil \textbf{Sub-6 GHz versus mmWave frequencies}\hfil}

\begin{figure}[!htbp]
\centering
\includegraphics[scale=.65]{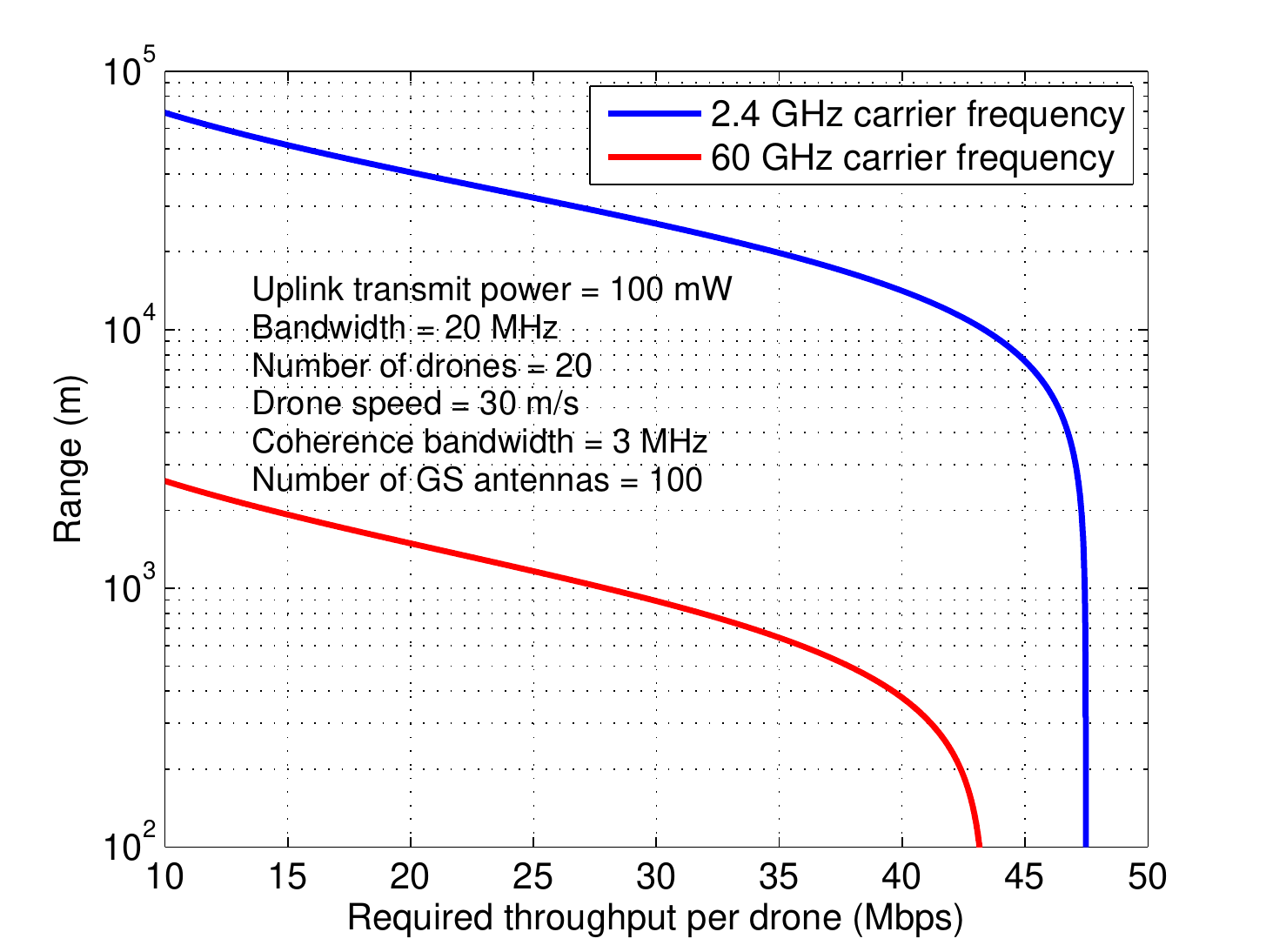}
\caption{Range vs. required throughput per drone at 2.4 GHz and 60 GHz carrier frequencies. The distribution of the drone positions and the orientation of the GS elements are the same as  in Figure \ref{Mag_sumThroughput_K}. Here,
for simplicity,  the  polarization loss is neglected.   ``Range'' refers to the coverage distance 
of the single antenna system for a given target data SNR. }
\label{Magazie_Sub6_mmWave_Comparison_upd}
\end{figure}

Massive MIMO can be used for drone communications at both sub-6 GHz and mmWave
frequencies. Here we make the following general observations:
\begin{itemize}
\item At mmWave frequencies, the energy efficiency is lower, due to decreased
  effective antenna areas. Suppose that the number of antennas and the target SNRs are fixed. Let $P_1$ and $P_2$ be the transmit
  powers required by a drone to maintain a target SNR at a particular distance with carrier frequencies $f_{c1}$ and $f_{c2}$, respectively. 
  From Friis's free-space equation, assuming unit antenna gains and
  perfect polarization match, the required transmit powers are related
  as \begin{equation}
	P_2= P_1 \left(\frac{f_{c2}}{f_{c1}}\right)^2. 
	\end{equation}
	For example, if
  $f_{c1}=2.4$ GHz and $f_{c2}=60$ GHz, then $P_{2}=625 P_1$.

\item The coverage at mmWave and sub-6 GHz frequencies will be significantly different as the coverage range is inversely proportional to the carrier frequency. From Friis's transmission formula, assuming unity antenna gains, for a given transmit power of $P_t$, bandwidth of $B$ Hz, and a target data SNR of $\rho_u$, the range is given by
\begin{equation}\label{range_eq}
	R= \frac{c}{4\pi f_c}\sqrt{\frac{P_t}{N_0 B \rho_u}} ,
	\end{equation}
	where $N_0$ is noise power spectral density in Watts/Hz.
\item Figure \ref{Magazie_Sub6_mmWave_Comparison_upd} shows the range 
supported at 2.4 and 60 GHz carrier frequency, respectively, for different throughput 
requirements with a 20 MHz bandwidth and a drone transmit power of 20 dBm. The range supported at 2.4 GHz is significantly higher than that at 60 GHz. For 
example, with a target throughput of 40 Mbps per drone, the supported range
 is 14 km and 370 m at 2.4 and 60 GHz, respectively. Note that here ``range'' refers to the coverage distance of the control channels. Since the control channel transmission takes place before and during the CSI acquisition, the range is the same as that of single antenna system and it remains constant irrespective of the number of GS antennas. In contrast, the range of data channels will depend on the number of GS antennas due to the coherent beamforming gain.
\begin{figure*}[!htbp]
\centering
\fbox{\subfigure[Disaster management during massive flooding]{\includegraphics[scale=.5625,trim={2.75cm 5.5cm 1cm 5.15cm},clip]{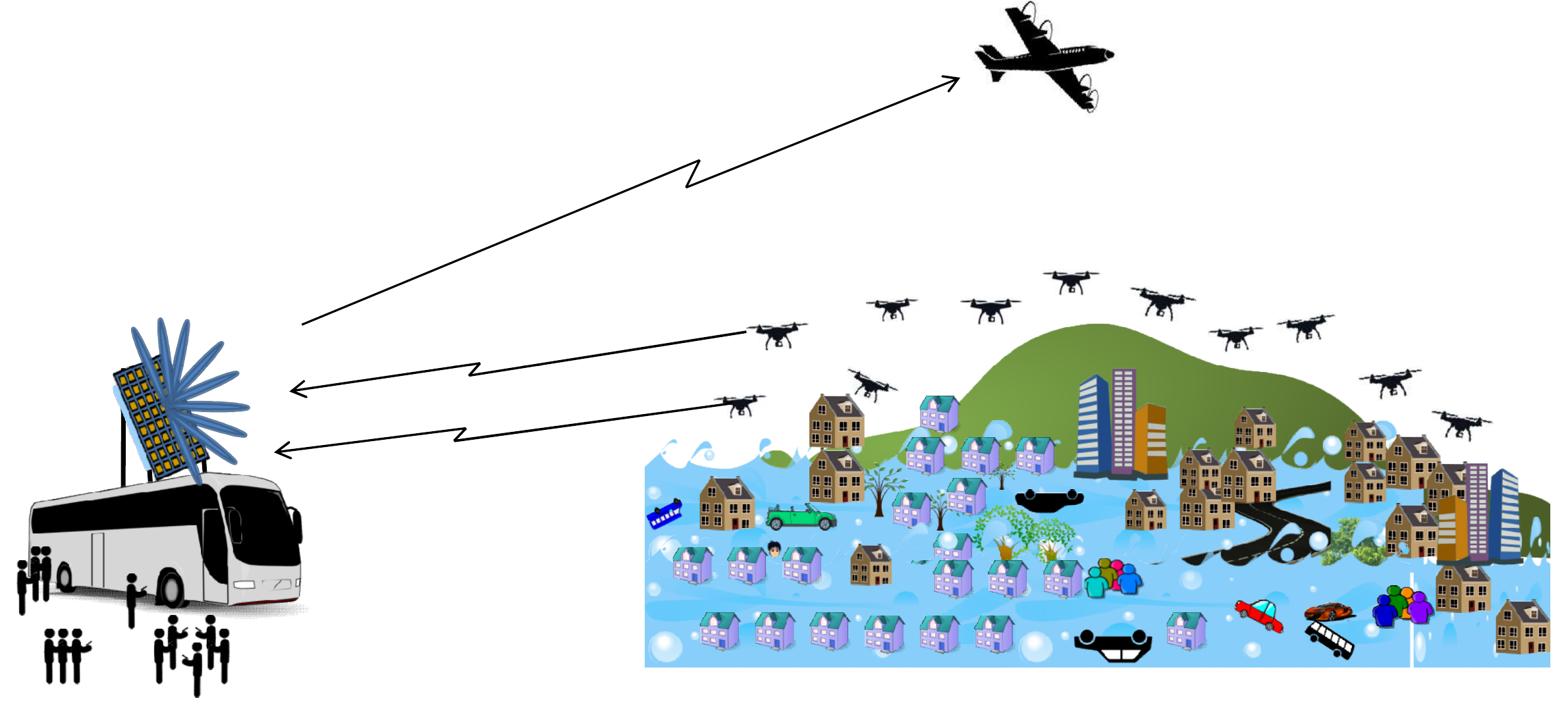}\label{Use-case1}}}\vspace{.1cm}
\fbox{\subfigure[Video-streaming at sport events]{\includegraphics[scale=.295,trim={1.5cm 2cm 3.5cm 5.75cm},clip]{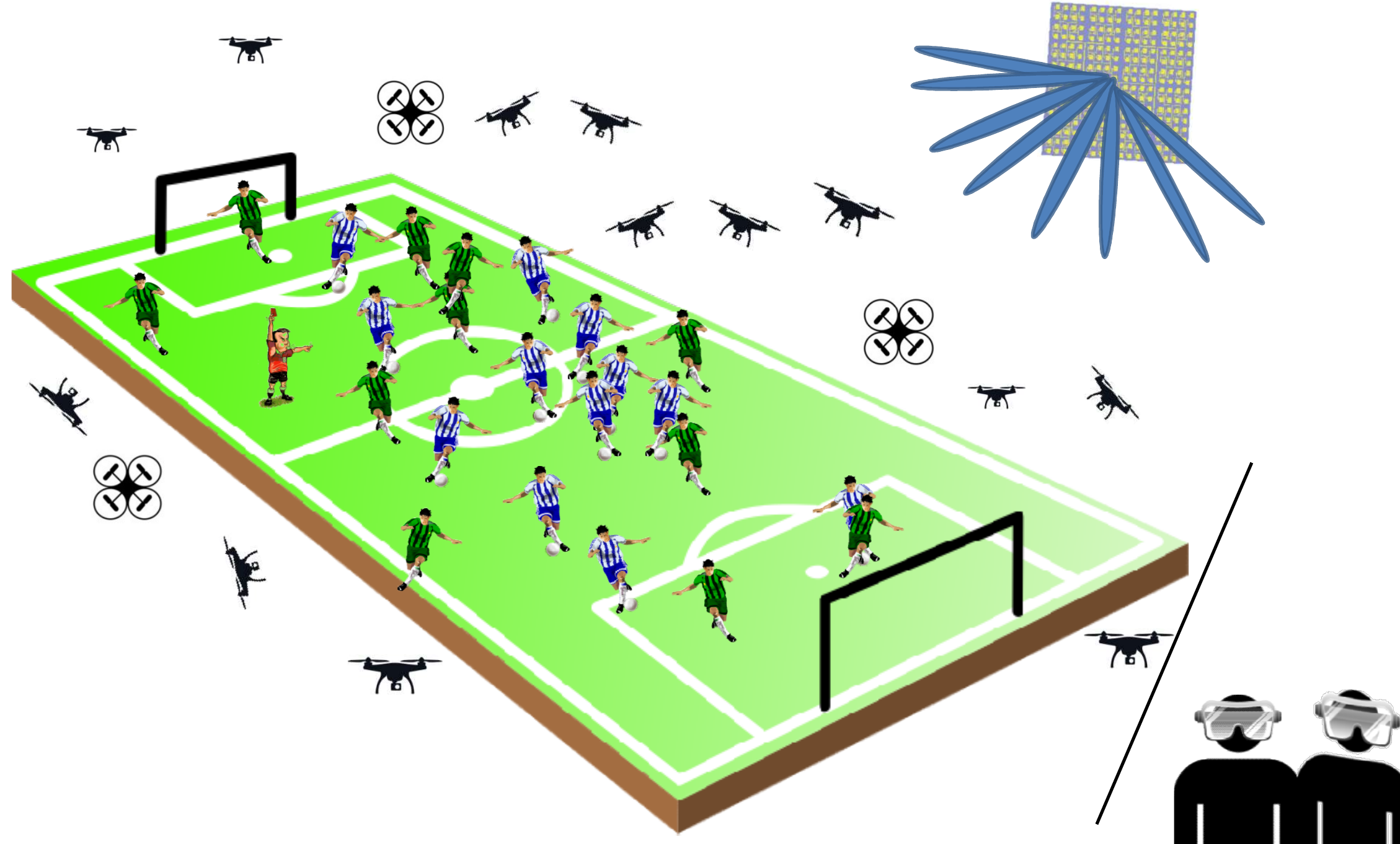}\label{Use-case2}}}
\fbox{\subfigure[Outdoor drone racing]{\includegraphics[scale=.25,trim={1cm 3.75cm 1cm 1.5cm},clip]{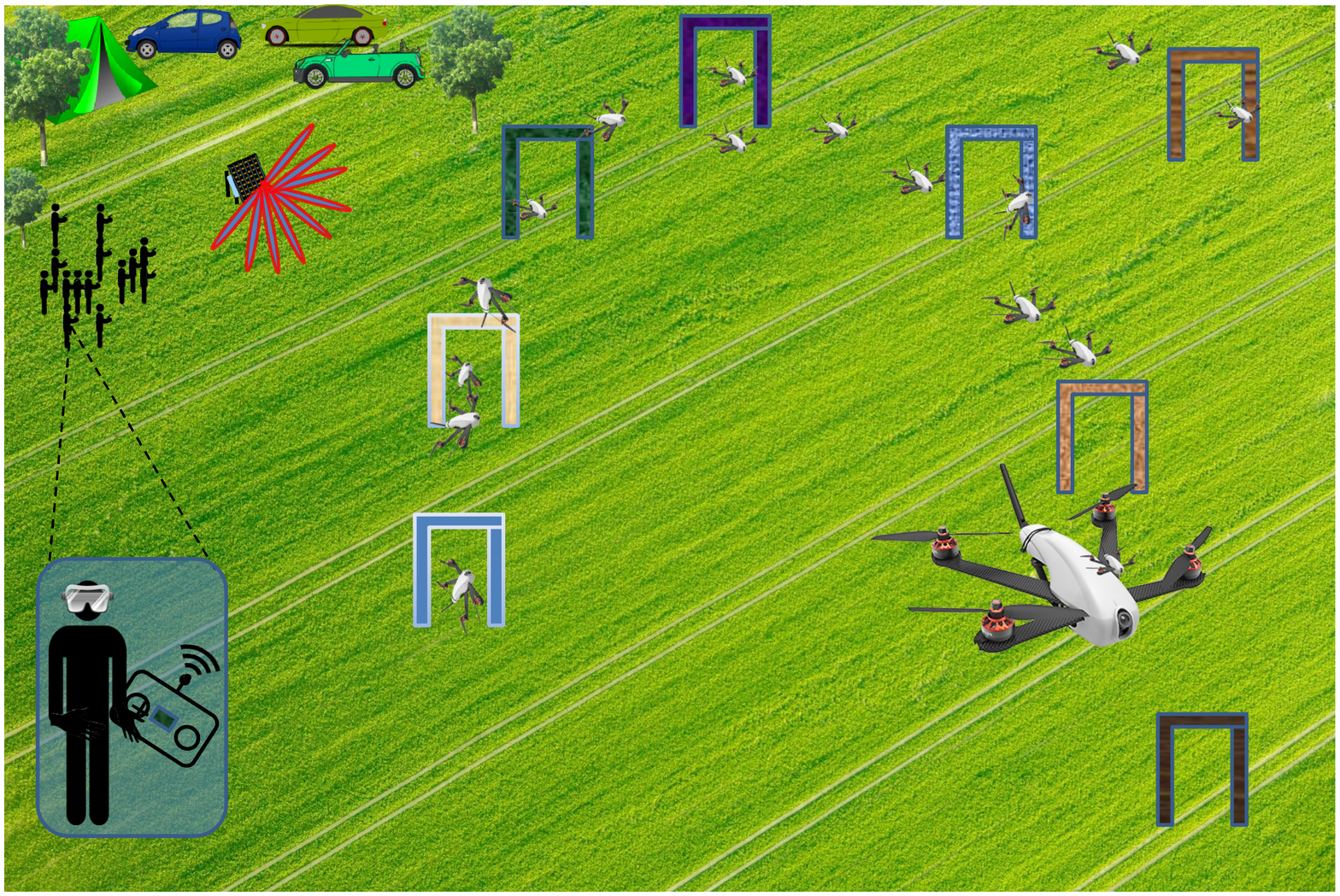}\label{Use-case3}}}
\caption{Massive MIMO use cases}
\label{use_cases}
\end{figure*}

\item The abundant bandwidth available at mmWave frequencies can be
  utilized for achieving high sum throughput (in the order of $10$
  Gbps). However, due to limitations on the drones' transmitted power, the
  coverage range might be limited. 
	
\end{itemize}

\subsection*{\hfil \textbf{Coverage extension removes the need for multi-hop solutions}\hfil}

Another important feature of Massive MIMO is the coverage extension
offered by the coherent beamforming gain. 
When compared with a single-antenna system, if the GS is equipped with $M$ antennas
 and under conditions when the GS can obtain accurate channel estimates,
   the coverage distance of the data channels can be extended
by a factor of $\sqrt{M}$. Thus Massive MIMO technology is a promising
alternative to multi-hop solutions, which suffer from issues with
reliability and latency, and difficulties in coordinating the
transmissions \cite{pinto2017}.

\subsection*{\hfil \textbf{Channel and interference hardening make low latency communication possible}\hfil}

In wireless links, variations in the channel lead to frequent
retransmissions. This often results in increased latency. But in
Massive MIMO, due to channel hardening, the effective channel gain
becomes deterministic (both in LoS and in fading) \cite{Marzetta16Book,narasimhan2014}. In multi-drone
networks, similar to channel hardening, \emph{interference hardening}
takes place with large numbers of antennas even in LoS propagation
conditions \cite{prabhu-spawc2017}. Therefore, an increased number of antennas at the GS
reduces the variations in the SINR. Stable SINR conditions help
satisfy low latency requirements.

\section{\textbf{Case studies}}
In this section we detail  three different case studies.

\subsection{\textbf{Case study I: Emergency response and disaster management}}

After natural disasters such as earthquakes or massive flooding, drone
swarms may be deployed rapidly and help rescue teams to assess the
situation in real time (see Figure \ref{Use-case1}). The video
received at the GS should have high quality, and in order to obtain
high-resolution imagery the drones have to fly at low altitudes. For
example, for a given dimension of the camera sensor (resolution ($r_{px}\times r_{py}$):
2664$\times$ 1496, focal length: 5$\times$ 10$^{-3}$ m and pixel size:
2.3$\times$ 10$^{-6}$ m), for achieving ground sampling distances (GSDs)
of 2, 20 cm, and 1 m, from Eqn. \eqref{altitude_label} we observe that, the required drone altitudes are
approximately equal to 44, 435, and 2174 m, respectively.

\paragraph*{\textbf{How many drones required for the mission?}}

Consider a geographical region to be covered with an area of 4 km
$\times$ 4 km, a drone camera resolution of 2664$\times$ 1496, and a
required GSD of 20 cm. The area covered by each camera image is
533 m $\times$ 300 m. With a single drone moving at 20 m/s, the
total time required to cover the area is more than seven hours. As the
flying time of small drones is often limited to 10--30 minutes, a
single-drone mission would additionally require a frequent return to
base. In contrast, a swarm of drones could cover the area in a short
time. For example, to complete the mission in 20 minutes, about 23
drones working in parallel are sufficient.

\begin{figure}[!htbp]
\centering
\includegraphics[scale=.65]{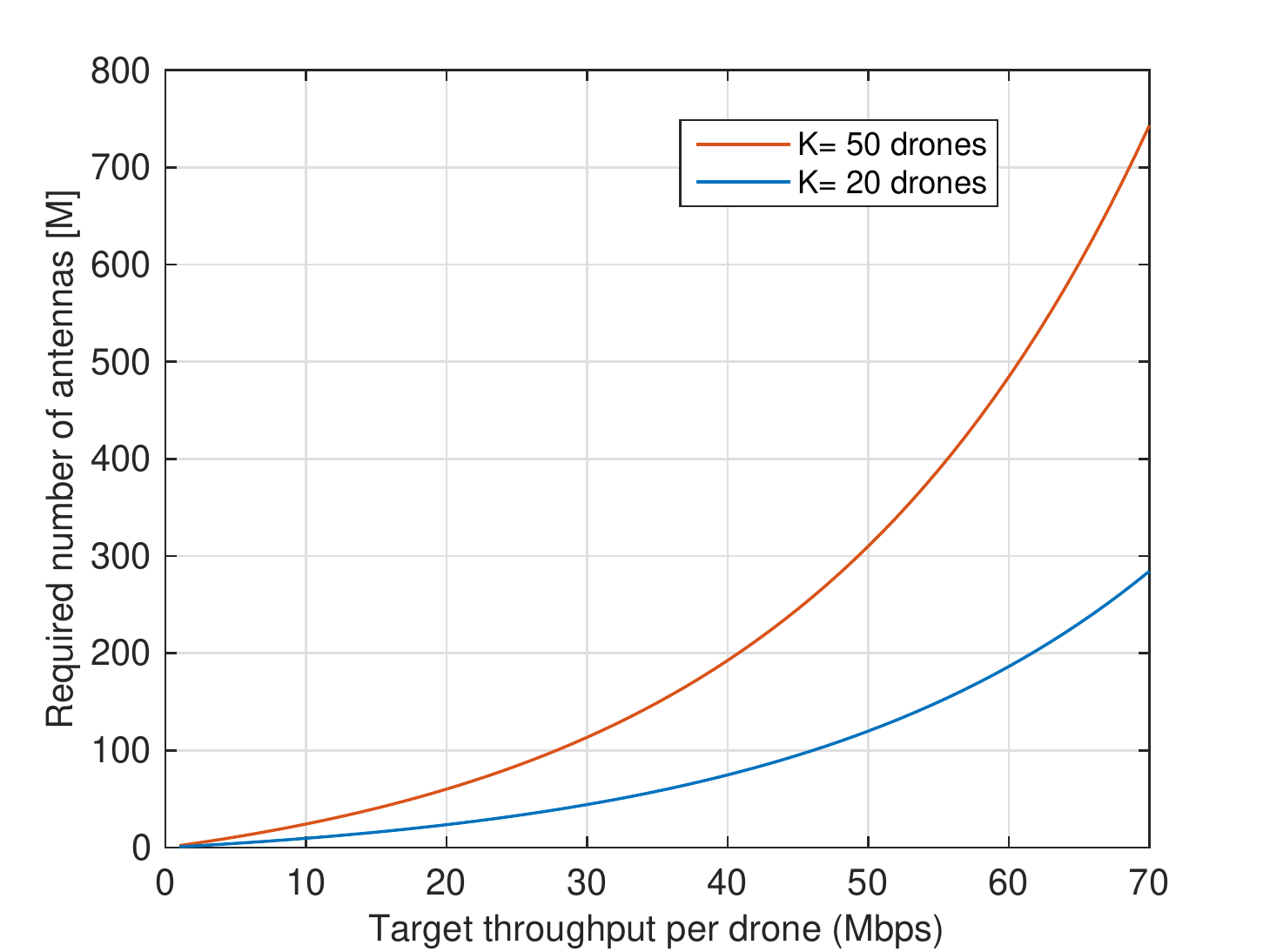}
\caption{Number of antennas required vs. target throughput per drone at 2.4 GHz carrier frequency for drone speed $v=$ 30 m/s, coherence bandwidth $B_c=$ 300 KHz, bandwidth $B=$ 20 MHz and SNR $\rho=$ 0 dB}
\label{M_req_vs_Target_rate}
\end{figure}

\begin{figure}[!htbp]
\centering
\includegraphics[scale=.65]{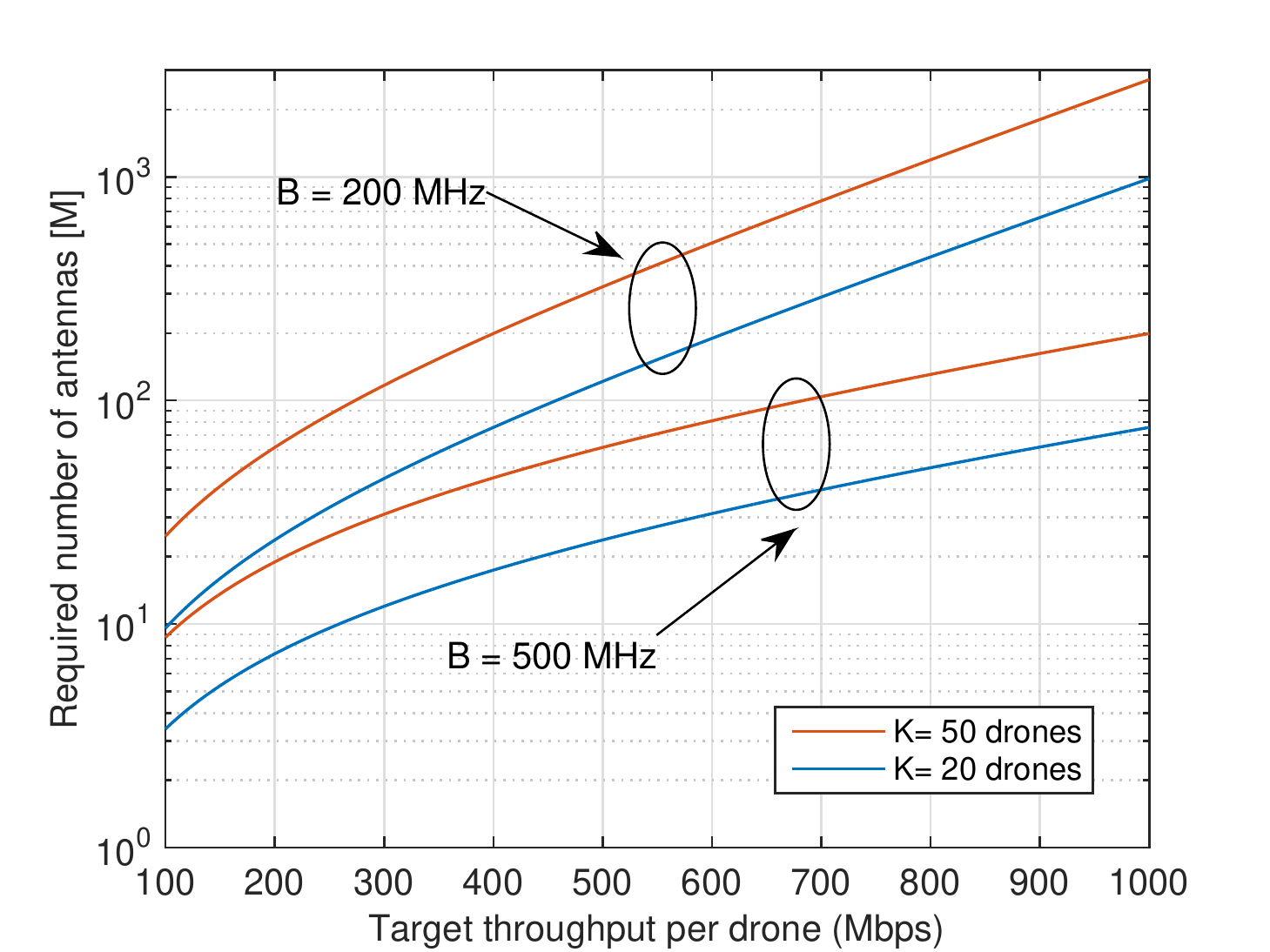}
\caption{Number of antennas required vs. target throughput per drone at 60 GHz carrier frequency for drone speed $v=$ 20 m/s, coherence bandwidth $B_c=$ 3 MHz and SNR $\rho=$ 0 dB}
\label{M_req_vs_Target_rate_mmWave}
\end{figure}

\paragraph*{\textbf{How much data need to be transmitted by a drone?}}

The uplink (drone to GS) data rate requirement depends on the desired
quality of the imagery. JPEG2000 is a loss-less compression standard
which gives compression ratios of up to 4:1. H264 and STD are
commonly used lossy video compression standards that achieve
compression ratios in the range 20:1 to 200:1. 
For video transmission, the throughput required by a drone is \cite{prabhu_TWC}
\begin{equation}\label{Q_video}
S_\mathrm{video}=\frac{r_{px}\cdot r_{py} \cdot b \cdot 
\mathrm{FPS} }{\mathrm{CR}} \ \text{(bits/sec)},\end{equation}
 where $b$ is the number of bits/pixel, $\mathrm{FPS}$  is the number of frames per second, and $\mathrm{CR}$ is the compression ratio. 
Table~\ref{case_study} gives the sum throughput requirement with a compression ratio of
200:1 and a GSD of 2 cm. A 4K resolution compressed video
requires the sum throughput of 1.46 Gbps. Depending on the
environment, more than 200 GS antennas are required to achieve this
throughput. The number of required antennas given in Table \ref{case_study} is obtained from Eqn. \eqref{asym_high_snr_mrc} as
\begin{align}
M_\mathrm{req} = &\left((K-1)+\frac{1}{\rho_u}+\frac{\kappa \chi_\mathrm{wc}}{\rho_u^2 \rho_p}(1+K\rho_u) \right) \nonumber\\
&\hspace{1cm}\times\left(2^{\frac{Q_\mathrm{tar}}{\left(1-\frac{2\cdot v\cdot f_c\left(K+\tau_{\mathrm{dl}}\right)}{B_c\cdot c} \right) B}}-1 \right)
\end{align}
and the range is obtained using \eqref{range_eq}.

\begin{table*}[!htbp]
\caption{Massive MIMO design parameters for different case studies }\label{case_study}
\tabulinesep=3mm
\hspace{-1.25cm}
    \begin{tabu}{ |  p{1.5cm}  | p{3.2cm} | p{3.7cm} | p{1.5cm} | p{3.25cm} | p{4.75cm} | }    \hline
    \hspace{-2.5cm}

    \textbf{Case study} &\begin{minipage}[t]{4cm}  \textbf{Drone parameters: \\ \textit{Number of drones ($K$),\\ Drone speed ($v$), \\Uplink transmit power ($P_t$)}} \end{minipage}&  \begin{minipage}[t]{3.7cm}\textbf{Camera parameters: \\ \textit{Pixel resolution (PR), \\Compression ratio (CR), \\Framers per second (FPS),\\ Angle of View (AoV), \\Number of bits per pixel (NBP)}} \end{minipage} &  \textbf{Required uplink sum Throughput ($K\times S$)} & \begin{minipage}[t]{5cm} \textbf{Channel parameters:\\ \textit{Coherence bandwidth ($B_c$),\\ Coherence time ($T_c$),\\ Coherence interval ($\tau$)}
}  \end{minipage}& \begin{minipage}[t]{4.7cm} \textbf{Massive MIMO design\\ parameters: \\\textit{
Carrier frequency ($f_c$), \\Bandwidth ($B$), \\Number of samples for downlink transmission ($\tau_{\mathrm{dl}}$),\\ Number of antennas ($M$), \\Data SNR ($\rho_u$),\\ Range ($R$) }
}  \end{minipage}\\ \hline
    \begin{minipage}[t]{2cm}\textbf{I. Disaster\\ management}   \end{minipage} &\begin{minipage}[t]{3cm} $K =$ 23\\ $v =$ 20 m/s \\
Altitude  = 100 m \\
$P_t = $100 mW\end{minipage}
 & \begin{minipage}[t]{3cm}PR = 4K\\
CR = 200:1 \\
FPS = 60\\
NBP = 24
 \end{minipage}& 1.39 Gbps&\begin{minipage}[t]{3.25cm}  $B_c$ = 3 MHz in sea and mountain environments (300 kHz in urban)\\
$T_{c}=3.125$ ms\\
$\tau=9375$ symbols (937 in urban)\end{minipage}
& \begin{minipage}[t]{4.5cm}
$f_c =$ 2.4 GHz, $B =$ 20 MHz\\
$\tau_{\mathrm{ul,d}}=\frac{9\tau}{10}$\\
$M =$ 216 (230 in urban)\\
$\rho_u =$ 0 dB \\
$R=$ 5 Km \end{minipage}
\\ \hline
	\begin{minipage}[t]{1.7cm}	\textbf{II. Sport \\ streaming }\end{minipage}        &\begin{minipage}[t]{3cm}  $K =$ 20\\ $v =$ 20 m/s\\
$P_t =$ 1 W 
 \end{minipage}
 &  \begin{minipage}[t]{3cm} PR = 4K 360 VR\\
CR = 200:1\\
FPS = 60\\
AoV = 90$^{\circ}$\\
NBP = 24 \end{minipage}

 & 19.3 Gbps &  \begin{minipage}[t]{3cm}$B_c =$ 3 MHz \\
$T_{c}=$ 0.125 ms\\$\tau=$ 375 symbols
\end{minipage}

 & \begin{minipage}[t]{5cm} $f_c =$ 60 GHz, $B =$ 300 MHz\\
$\tau_{\mathrm{ul,d}}=\frac{9\tau}{10}$\\
$M =$ 260 (1035 with $B =$ 200 MHz)\\
$\rho_u=$ 0 dB \\
$R=$ 160 m (200 m with $B =$ 200 MHz)\end{minipage}

\\ \hline

	\begin{minipage}[t]{3cm}\textbf{ III.	Drone\\ racing }  \end{minipage}
       & \begin{minipage}[t]{3cm}$K = 25$\\ $v = 30$ m/s \\
$P_t =100$ mW \end{minipage}

 &  \begin{minipage}[t]{3cm} PR = $640\times 480$\\
CR = $1:1$\\
FPS = $30$\\
AoV = $120^{\circ}$\\
NBP = $8$
 \end{minipage}
& 1.84 Gbps & \begin{minipage}[t]{3.5cm}$B_c =$ 3 MHz\\
$T_{c}=$ 0.862 ms\\
$\tau=$ 2586 symbols\end{minipage}
& \begin{minipage}[t]{5cm} $f_c =$ 5.8 GHz, $B =$ 20 MHz\\
$\tau_{\mathrm{ul,d}}=\frac{9\tau}{10}$\\
$M =$ 420 (840 for $K =$ 50)\\
$\rho_u=$ 0 dB \\
$R=$ 2 Km \vspace{.2cm}
 \end{minipage}
\\ \hline

    \end{tabu}
		\end{table*}

At 2.4 GHz carrier frequency, Figure \ref{M_req_vs_Target_rate} illustrates the number of antennas required for a given data rate per drone with 20 and 50 drones. At 60 GHz carrier frequency, Figure \ref{M_req_vs_Target_rate_mmWave} shows the number of antennas required for a given data rate per drone with 200 MHz and 500 MHz bandwidths for 20 and 50 drones.

\subsection{\textbf{Case study II: Real-time video streaming at sports events}}

At large-scale sports events, 20 or more quadcopters equipped with
4K (4096 $\times$ 2048) resolution 360-degree cameras can be used
to capture the players' actions at multiple angles (see Figure
\ref{Use-case2}) \cite{wang2017_spst}. The captured videos can be sent
to an access point equipped with a massive antenna array. This enables
real-time virtual reality for the viewers. However, it requires a very
high throughput of the wireless link, depending on the desired AoV. For example, using a head-mounted display, with a
horizontal AoV of 90 degrees, at any given moment the viewer can see
only one-fourth of the scene (or one-third of the scene with a
120-degree AoV).\footnotemark Thus with 4K resolution and
90-degree AoV, each frame will have 16384$\times$ 8192 pixels. With
60 frames per second and 24 bits/pixel, every camera produces
approximately 184 Gigabits per second. With a 200:1 compression
ratio, the bit rate required by the wireless link is 920 Mbps. If
there are 20 drones covering the event, the required sum
throughput is 18.4 Gbps. \footnotetext[1]{A 360 video features a complete panoramic 360-degree horizontal view and a 180-degree vertical view.}

A typical radius of a large football stadium is approximately 230
m. If we consider that the antenna array is placed on the rooftop of
the stadium, the distance between the array and drones will be between
50--100 m. Due to this short range, Massive MIMO technology at
mmWave frequencies is a suitable choice. As shown in Table
\ref{case_study}, at 60 GHz frequency, over 300 MHz bandwidth,
with a drone transmit power of 1 W, it is possible to achieve 0 dB
SNR at 160 m distance (50 m with 100 mW transmit power). For 20 drones, the number of antennas
required to achieve the sum throughput of 18.4 Gbps is 227.

\subsection{\textbf{Case study III: Drone Racing}}

Drone racing, often referred as ``the sport of the future,'' is
attracting big sponsorship deals and venues all over the world
\cite{Northfield2013,Schneider2015}. In drone racing contests, see Figure
\ref{Use-case3}, pilots wearing goggles control the movement of
quadcopters via a First Person View (FPV) interface. The drones can fly at speeds exceeding 100 km/h.  Control of the drone movement requires very
low latency (tens of milliseconds). Currently, analog transmission is
used to reduce the delay between the video capture and display. Due to
limited capacity, the maximum number of drones in a race is also
limited to 8. Using a massive GS antenna array, hundreds of drones
could be simultaneously served on the same time-frequency resource
with low latency.

\paragraph*{\textbf{What is the maximum acceptable latency for drone control and video transmission?}} 

For video transmission, latency represents the time difference between
the instant a frame is captured by the drone and the instant that
frame is shown on the pilot's display. Including the video capture
time, and coding- and decoding delays, the overall end-to-end latency
is more than 120 ms \cite{TI_Latency}. Specifically, standard
compression techniques such as H264, STD introduce about 50 ms
latency. This delay has serious impact as controlling a drone's
movement in a 3D space requires very low latency. For example, at
40 m/s, a quadcopter moves 2 meters during 50 ms. Adding this
lag to the control loop for the pilot makes the flight extra
challenging, especially in indoor environments.

The efficiency achieved by standard video compression schemes
depends on the differences between subsequent frames. Thus, the
successful delivery of a frame depends on the previous frames. As a result, the higher the compression, the greater the possibility of a frame error and frequent retransmissions increase latency. In such situations, one approach to reduce the end-to-end latency is to transmit uncompressed videos by avoiding compression and decompression modules. The end-to-end latency can be made negligible by the use of Massive MIMO enabled high throughput wireless link. To simultaneously support 25 drones in a racing contest, with a
moderate video quality, the sum throughput required for transmitting
raw video is 1.84 Gbps (with latency less than 70 ms). As shown in
Table \ref{case_study} about 415 antennas are required to achieve
1.84 Gbps sum throughput.

\begin{table*}[!htbp]
\caption{Air-to-ground channel measurement results}
\label{air_ground_table}
\hspace{-.875cm}
    \begin{tabular}{|l|l|l|l|l|l|l|}    
    \hline
	    \textbf{References} &  \textbf{Scenario} &  \textbf{Altitude} &  \textbf{Antenna polarization} & \textbf{Frequency} & \textbf{No. of antennas} & \textbf{Measurement parameters}\\ 
    \hline
    
    	   \cite{meng2011}	&Over-sea & 370 m - 1.83 km & Vertical& 5.7 GHz & 1 Tx - 1 Rx  & Pathloss \& RMS Delay spread \\
	\hline
	   \cite{matolak2017_water}	&Over-sea & 800 m & Vertical&968 MHz \& 5.06 GHz & 1 Tx - 1 Rx  & Pathloss \& RMS Delay spread \\
	\hline
    \cite{sun2017_hilly} & Hilly \& Mountain  & 1 km - 4 km & Vertical& 968 MHz \& 5.06 GHz & 1 Tx - 1 Rx  &Pathloss \& RMS Delay spread  
    \\ 
    \hline
        \cite{matolak2017_urban} & Urban  & 750 m - 1 km & Vertical& 968 MHz \& 5.06 GHz & 1 Tx - 1 Rx  & Pathloss \& RMS Delay spread 
    \\ 
    \hline
        \cite{sun2017_airframe} & Desert  & 900 m - 1 km& Vertical& 968 MHz \& 5.06 GHz & 1 Tx - 1 Rx  & Air-frame shadowing loss
    \\ 
    \hline
		\cite{kovacs2017,amorim2018_wcl}& 
	Rural & 1.5 m - 120 m &Vertical& 800 MHz& 1 Tx - 1 Rx  &Signal-to-interference ratio\\ 
	\hline
	\cite{yanmaz2013} & Outdoor  & 100 m & Horizontal \& Vertical &5 GHz & 1 Tx - 1 Rx  & Received signal strength\\ 
	\hline
	\cite{newhall2003} & Urban  & 450 m - 1 km & Vertical& 2 GHz &1 Tx - 4 Rx & RMS Delay spread\\ 
	\hline
		\cite{qualcomm_lte_trial} & Sub-urban  & 30 m - 120 m & N/A & 700 MHz \& 1.9 GHz & N/A & Received signal strength\\ 
	\hline
			\cite{guangyang2018} & Urban  & 50 m - 300 m & N/A &2.6 GHz & N/A & Latency \& Received signal strength\\ 
	\hline
				\cite{lin2018_archive} & Urban  & 50 m - 150 m & N/A &800 MHz & N/A & Received signal strength\\ 
	\hline
					\cite{alhourani2018} & Sub-urban  & 15 m - 120 m & N/A &800 MHz & N/A & Received signal strength\\ 
	\hline	
						\cite{asadpour2014} & Indoor  & N/A & Vertical \& Circular&5 GHz & N/A & Received signal strength\\ 
	\hline	
	\end{tabular}
\end{table*}

\section{\textbf{Future Research Directions}}

A Massive MIMO-enabled drone communication system design involves many challenges. In this section we detail some of the potential research issues that need to be addressed in order to fully exploit  the Massive MIMO technology.

\subsection*{\hfil \textbf{Scheduling, resource optimization and 3D trajectory optimization}\hfil}

    In practice, there are situations where the channel is changing very slowly (for example, consider two hovering drones at high altitudes with the same azimuth and elevation angles) or the drones positions are closely located \cite{Kushleyev2013}. In such conditions the antenna array at the GS might not be able to spatially resolve the channels of the drones and spatial multiplexing becomes impossible; the GS will have to schedule the interfering drones on different time-frequency resources \cite{Yang2018}. An alternative approach, if bandwidth is limited and throughput requirements are high, could be to jointly plan the trajectories and the scheduling \cite{bulut2018}.

\subsection*{\hfil \textbf{New MAC layer design}\hfil}
With Massive MIMO, the frequency of the effective interference variation depends on the GS array geometry, the flying speed and altitude, and range. Furthermore, if the mobility patterns of the drones are known, frequent pilot transmission for CSI estimation is not necessary. Also in many applications, uplink traffic is much higher than the downlink traffic. Thus, a flexible TDD frame structure may be designed taking into account these factors \cite{pedersen2015,si2015}.

New flexible MAC layer designs are possible, because unlike existing wireless standards, drone networks do not have backward-compatibility requirements. For video and image transmissions, along with Massive MIMO, efficient cross-layer designs can be utilized to enhance the performance of drone networks \cite{immich2014,karzand2017,ilter2018}.

\subsection*{\hfil \textbf{Air-to-ground channel  models}\hfil}
The propagation conditions in aerial communications are different from terrestrial land-mobile radio environments. Especially, LoS conditions are generally prevalent with Rician K-factors typically exceeding 25 dB \cite{matolak2017_water,sun2017_hilly,matolak2017_urban,amorim2017_wcl,alshbatat2010}. Table \ref{air_ground_table} shows a list of air-to-ground channel measurement results carried out in the recent years. For more detailed survey on air-to-ground channel models, see \cite{khawaja2018_ch_model_survey}. The root-mean-square (RMS) delay spread of the channel is an important parameter for Massive MIMO based system design as it determines the coherence interval. For example, measurement results in the C-band (5.03--5.091 GHz), show that the average RMS delay spread is on the order of 10 ns (with an average UAV altitude of 600 m and link ranges from 860 m to several kilometers) in over-water, hilly and mountainous terrain, and 10--60 ns in suburban and near-urban environments \cite{matolak2017_urban}.  Hence, the coherence bandwidth, if defined as the frequency separation associated with a correlation of 0.5 \cite{sesia2009}, varies between 3 MHz and 20 MHz.

Although measurement results are unavailable at mmWave frequencies
(30 GHz to 300 GHz), due to the reduced number of multipath
components, one can expect even higher coherence bandwidths than at
sub-6 GHz frequencies.

Currently, the measurement results are available only for linearly polarized (horizontal or vertical) antennas. Moreover, the measurements have been taken without altering the orientation of drone antennas. Measurement campaigns need to be carried out, preferably using circularly polarized antennas, over a wide range of carrier frequencies in various environments such as over-water, mountainous, and urban.

\subsection*{\hfil \textbf{Spectrum management}\hfil}

A drone network's bandwidth requirement depends on several factors,
such as the environment, communication range, density of drones,
carrier frequency, duplex mode, power source, and throughput
requirements \cite{Kakar2017,si2015}. In applications with low drone
densities, the spectrum can be opportunistically shared with existing
cellular and satellite communication systems. However, dedicated
spectrum may be required in case of control and non-
payload communications (CNPC) and applications that involve high drone densities \cite{kerczewski2013}. Furthermore,
unlicensed spectrum may be utilized \cite{dronealliance_eu}.  The spatial multiplexing
capabilities of Massive MIMO will be instrumental in reducing the
spectrum requirements.

\textit{Which duplexing mode: FDD or TDD?} TDD is generally more
efficient in reciprocity-based MIMO systems \cite{Marzetta16Book,vieira2017}. In
FDD mode, the number of samples required for CSI acquisition is the
limiting factor. The number of uplink pilot symbols per coherence
interval is at least equal to the number of drones in TDD. On the other hand, in FDD, it is
 at least equal to the   number of GS antennas plus the number of drones. However, due to the
reduced number of multipath components in drone communication
scenarios, beam tracking may be feasible, which would reduce the need
for CSI acquisition.  Detailed analysis is required of different
duplexing modes in different environments and applications.

\subsection*{\hfil \textbf{Networking challenges}\hfil}
Future aerial networks will be heterogeneous, and consist of different types of drones: high-altitude and long-endurance UAVs, medium-range UAVs, short-range UAVs, and mini-drones \cite{yanmaz2013,zeng2016,wang2017,si2015}. Nodes in various parts of the network may move at different speeds and have different numbers of antennas. For example, in some applications, at mmWave frequencies, the drones themselves can be equipped with antenna arrays to achieve additional capacity gains \cite{zguan2018,cuvelier2018,hu2016}. In other applications, the antenna elements may be distributed over a large geographical region \cite{she2019}. Massive MIMO technology in these cases poses new requirements on synchronization.

Moreover, if transmissions from different GSs are uncoordinated, the SINR may fluctuate unpredictably and create highly non-stationary interference, known as flashlight interference \cite{ihalainen2013}. Suitable interference management techniques are required for efficient operation of Massive MIMO based drone networks with multiple GSs.

\section{\textbf{Summary}}
Massive MIMO technology is an enabler for wireless connectivity with swarms of drones in applications which require long-range and high-throughput links. This entails equipping the GS with a large antenna array, whereas each drone can have a single antenna. Compelling features of Massive MIMO for this application include: high throughput, spatial multiplexing, coverage extension and high-mobility support. We identified several new and important research problems that are relevant to future Massive MIMO-enabled drone communication networks.

\end{document}